\documentclass[a4paper,nofootinbib,useAMS,usenatbib]{mn2e}
\usepackage{amssymb}
\usepackage{amsmath}
\usepackage{epsfig}
\usepackage{subfigure}
\usepackage{mathrsfs}
\usepackage{longtable}
\usepackage[usenames,dvipsnames]{xcolor}
\usepackage{mathtools}
\usepackage{relsize}
\usepackage{geometry}
\usepackage{graphicx}
\usepackage{bm}
\usepackage{caption}

\usepackage{color}

\makeatletter
\newcommand*{\rom}[1]{\expandafter\@slowromancap\romannumeral #1@}
\makeatother

\begin{document}
\newcommand{\tkDM}[1]{\textcolor{red}{#1}}  
\newcommand{\tkBB}[1]{\textcolor{blue}{#1}}  
\newcommand{\tkRW}[1]{\textcolor{ForestGreen}{#1}}  
  \newcommand{\JS}[1]{\textcolor{brown}{#1}}  
\newcommand{\unit}[1]{\ensuremath{\, \mathrm{#1}}}
\newcommand{\angstrom}{\mbox{\normalfont\AA}}

\title[Constraints on Axion Dark Matter]{Galaxy UV-luminosity function and reionization constraints on axion dark matter}

\author[B. Bozek et al] {Brandon Bozek$^{1,2}$\thanks{bbozek@astro.umd.edu}, David~J.~E.~Marsh$^3$, Joseph~Silk$^{2,4,5}$, Rosemary F.G.~Wyse$^2$.\\
$^1$Department of Astronomy and Joint Space-Science Institute, University of Maryland, College Park, MD 20742, USA \\
$^2$Department of Physics and Astronomy, The Johns Hopkins University, Homewood Campus, Baltimore MD 21218, USA \\
$^3$Perimeter Institute, 31 Caroline St N,  Waterloo, ON, N2L 6B9, Canada\\ 
$^4$Institut dÕAstrophysique, UMR 7095 CNRS, Universit\'{e} Pierre et Marie Curie, 98bis Blvd Arago, 75014 Paris, France \\
$^5$Beecroft Institute of Particle Astrophysics and Cosmology, Department of Physics, University of Oxford, Oxford OX1 3RH, UK
}
\date{\today}
\maketitle

\begin{abstract} 

If the dark matter (DM) were composed of axions, then structure
formation in the Universe would be suppressed below the axion
Jeans scale. Using an analytic model for the halo mass function of a
mixed DM model with axions and cold dark matter, combined with the
abundance-matching technique, we construct the UV-luminosity
function. Axions suppress high-$z$ galaxy formation and the
UV-luminosity function is truncated at a faintest limiting
magnitude. From the UV-luminosity function, we predict the
reionization history of the universe and find that axion DM causes reionization to occur at lower redshift. We search for evidence
of axions using the \textit{Hubble Ultra Deep Field} UV-luminosity function in
the redshift range $z=6$--$10$, and the optical depth to
reionization, $\tau$, as measured from cosmic microwave background polarization. All probes we
consider consistently exclude $m_a\lesssim 10^{-23}\text{ eV}$ from
contributing more than half of the DM, with our strongest constraint
ruling this model out at more than $8\sigma$ significance. In
conservative models of reionization a dominant component of DM with
$m_a=10^{-22}\text{ eV}$ is in $3\sigma$ tension with the measured
value of $\tau$, putting pressure on an axion solution to the
cusp-core problem. Tension is reduced to $2\sigma$ for the axion contributing 
only half of the DM. A future measurement of the UV-luminosity
function in the range $z=10$--$13$ by \textit{JWST} would provide further
evidence for or against $m_a=10^{-22}\text{ eV}$. Probing still higher
masses of $m_a=10^{-21}\text{ eV}$ will be possible using future
measurements of the kinetic Sunyaev--Zel'dovich effect by Advanced ACTPol to
constrain the time and duration of reionization.

\end{abstract}


\section{\label{sec:intro} Introduction}

While dark matter (DM) is known to comprise a large portion of the
energy density of the universe, $\Omega_{\rm d} h^2\approx 0.12$ \citep[e.g.][]{Ade:2013zuv}, and plays an
important role in the formation and dynamics of galaxies and clusters, its particle nature is unknown. Two leading
candidates in well-motivated and minimal extensions of the standard
model of particle physics (the SM) are weakly-interacting massive
particles (WIMPs), which emerge naturally in
supersymmetry \citep{Jungman:1995df}, and the QCD axion
\citep{pecceiquinn1977,weinberg1978,wilczek1978}, which solves the CP
problem of strong interactions. Many experimental efforts are
underway to detect and constrain these DM candidates via their direct \citep[e.g.][]{,Asztalos:2009yp,Aalseth:2010vx,Aprile:2012nq,Agnese:2014aze,Akerib:2013tjd,Angloher:2014dua,Budker:2013hfa,2014arXiv1409.2986S} or
indirect \citep[e.g][]{Brockway:1996yr,Grifols:1996id,Ackermann:2011wa,IceCube:2011aj,Aguilar:2013qda,Friedland:2012hj,Blum:2014vsa} interactions with the SM, but no definitive
evidence has so far emerged.\footnote{The particle physics status of DM, including LHC searches for supersymmetric WIMPs, is reviewed in \cite{Beringer:1900zz}. For reviews of axion physics see \cite{raffelt2001} and \cite{Wantz:2009it}. Of course supersymmetry and axions are not mutually exclusive: indeed they are necessary partners in string theory \citep{witten1984,witten2006}. For a review of DM models with supersymmetric axions see \cite{Baer:2014eja}.} 

As these experiments designed to directly detect WIMPs and axions continue
to report null results \citep{Beringer:1900zz,Akerib:2013tjd}, with associated shrinkage
of allowed parameter space, we are motivated to try to constrain the
particle nature of DM via the only interaction it is known to have:
gravitation. Further, we will explore models beyond those where WIMPs and axions
comprise all of the DM and consider scenarios where the DM is multi-component. 

We can go further than measuring the DM density and can constrain the
physics of DM should it affect the formation and growth of structure
in a novel way. Cold (C)DM clusters on all scales and makes
well-understood predictions relating to the formation and growth of
cosmic structure \citep{1971phco.book.....P,1983ApJ...274..443B,1984Natur.311..517B,1985ApJ...292..371D}. Standard supersymmetric WIMPs have $\mathcal{O}(\text{GeV})$
masses and are thermally produced, leading to negligible
free-streaming lengths -- the defining characteristic of
cold dark matter (CDM). The QCD axion is much lighter than a WIMP, with
$\mathcal{O}(\mu\text{eV})$ mass but, since it is non-thermally
produced, it too is gravitationally indistinguishable from CDM due to
vanishing sound-speed \citep[e.g.][]{noh2013}. Since both standard WIMPs and
QCD axions are equivalent to CDM in structure formation,\footnote{If
axion DM were to form a Bose-Einstein condensate, then some features
such as vortices or caustics in galaxies may occur \cite[e.g.][]{sikivie2010b}.}
in order to learn about the particle nature of DM via gravitational
probes it must cluster in a manner distinct from CDM.  Constraining
the particle nature of DM using the growth of structure therefore
requires considering models other than standard WIMPs
and the QCD axion.

\section{Probing the nature of DM using structure formation}
\subsection{Models, motivations, and existing bounds}

Two popular models that manifest novel structure formation are warm
(W)DM \citep[e.g][]{1982PhRvL..48.1636B,bode2001} and ultralight axions \citep[ULAs,
e.g.][]{axiverse2009}. The low-mass scale necessary for ULAs can be naturally realized in string theory models \citep[e.g.][]{acharya2010a,cicoli2012c}, or in the hidden-sector model of \cite{chiueh2014}. Alternatively it could occur for the QCD axion with an extremely super-Planckian decay constant, but this is considered theoretically problematic \citep[e.g.][]{arkani-hamed2007}.

Each of these models has additional motivation
since both WDM and ULAs suppress small-scale structure and can
help in the resolution of the small-scale problems of CDM, which
include the over-prediction of low-mass dark haloes \citep[`missing satellites'; ][]{1999ApJ...524L..19M,1999ApJ...522...82K}; the prediction of a central `cusp' in
the DM density profile while observations favour `cores' \citep{2008IAUS..244...44W}; the
prediction of more numerous satellite galaxies of the mass of the
Large Magellanic Clouds \citep[the `too-big-too-fail' problem; ][]{2011MNRAS.415L..40B} and the
difficulty in producing typical disc galaxies due to the prediction of
active mergers until redshift of order unity \citep{2001ASPC..230...71W}. 

 The adoption of a lower mass thermally produced DM particle, such as in the  WDM scenario, introduces a tension between the desire to produce a core in the inner regions of the DM density profile -- favouring a lower mass, while simultaneously producing dwarf galaxies in sufficient
 number -- favouring a higher mass 
 \citep{maccio2012,schneider2013b}. Other effects of WDM, in the case of an $\sim 1 \rm  keV$ neutrino, include a significant impact on faint galaxy counts \citep{2014MNRAS.442.1597S} and early star formation rates \citep{ 2014arXiv1408.1102D}.
 
 Following the work of
 \cite{hu2000}, it was shown in \cite{marsh2013b} that ULAs are not subject to this tension due to the inverse relationship between halo mass and
 core size in these models. Recently, high-resolution simulations of
 core formation with ULAs by \cite{schive2014,schive2014b} have confirmed this
 picture. Fits to the cored halo profile in Fornax give a best-fitting
 mass of $m_a=8.1^{+1.6}_{-1.7}\times 10^{-23}\text{ eV}$, providing a large core while still forming low-mass galaxies. 

Cosmological probes of the linear regime of
structure formation, such as the power spectrum, $P(k)$, of density fluctuations \citep[e.g.][]{reid2010} and the cosmic microwave background \citep[CMB; e.g.][]{Ade:2013ktc}, provide only weak constraints on the mass of
the DM particle (warm or axion-like) in each of these scenarios.\footnote{From structure formation it is well established that  that the DM is not `hot', for example composed of light neutrinos \citep{1979PhRvL..42..407T,1982PhRvL..48.1636B,White:1984yj}. For  DM  composed of CDM plus massive neutrinos, the CMB limits the total neutrino mass as $\sum m_\nu<0.66\text{ eV}$ at 95 per cent C.L. \citep{Ade:2013zuv} and neutrinos contribute a sub-per cent fraction of DM.}. The WDM cannot be
`too warm', for example we must have $m_W\gtrsim 0.1\text{ keV}$, and ULAs cannot be too light either,
$m_a\gtrsim 10^{-24}\text{ eV}$ \citep[Marsh et al., in preparation]{amendola2005}. However, since one requires $m_W\sim \mathcal{O}(1)\text{ keV}$ and $m_a\sim 10^{-22}\text{ eV}$ in order for WDM or ULAs to be relevant to the small-scale problems, linear probes are silent on the validity of these scenarios. 

Thus, in order to test and constrain a particle physics solution to the
small-scale problems, we must look to non-linear probes of structure
formation. One such probe is the Ly$\alpha$ forest flux power
spectrum. \cite{viel2013} have used observations of the Ly$\alpha$ forest, combined with hydrodynamic simulations of structure formation, to place the
strongest constraint to date on WDM, $m_W>3.3\text{
keV}$. \cite{amendola2005} used older Ly$\alpha$ data to constrain
ULAs, placing the bound $m_a>5\times 10^{-23}\text{ eV}$ if the ULA is
to be all of the DM. Analyses of the Ly$\alpha$ forest involve considerable complexity 
related to the non-linear mapping of the optical depth 
and to the required calibration from simulations involving gas physics. No
detailed predictions have been made for the Ly$\alpha$ forest with ULAs as DM, as such
simulations do not exist. Existing Ly$\alpha$ constraints on WDM and ULAs
point to larger masses of the scale we hope to constrain, and if
properly understood will be able to provide consistency and cross-checks.\footnote{Another probe that can constrain non-linear scales is
galaxy weak lensing, for example through the anticipated data sets from \textit{Euclid}
\citep{Laureijs:2011gra,Amendola:2012ys}. \cite{smith2011} forecast that \textit{Euclid} may be able to
constrain $m_W\gtrsim 2.6 \text{ keV}$, and prospects for ULAs also
look promising \citep[Marsh, in preparation]{marsh2011b}, though
considerable experimental and theoretical systematics are
involved.}

\subsection{The UV-luminosity function and reionization}

The two probes we will focus on in this work are one, 
the UV-luminosity function of galaxies at high-redshift, $\phi (z)$, as measured by the \textit{Hubble Space Telescope} e.g.~\cite{2014arXiv1403.4295B} and two, the
reionization history of the universe through the Thomson scattering optical depth to reionization, $\tau$, measured from large-angle CMB polarization by \textit{WMAP} \citep{Bennett:2012zja}.\footnote{We use the $\tau$ likelihood derived from \textit{Planck}+\textit{WMAP} chains in \cite{spergel2013}.}

The extremely deep imaging in bandpasses from the optical to near-IR with the \textit{Hubble Space Telescope} (\textit{HST}) available in several Legacy Fields provides the most fundamental data set for
constraining the contribution of galaxies to the reionization of the
universe through estimation of the rest-frame UV-luminosity function over the redshift range between
$z=4$ and $10$ \citep{2011ApJ...737...90B,2014arXiv1403.4295B,2012ApJ...759..135O,2013MNRAS.429..150L,2013MNRAS.432.2696M}. The 
rest-frame UV-luminosity function is a measure of  the number density, per absolute magnitude, of the star-forming galaxies that are likely to be the primary driver of
reionization \citep{1999ApJ...514..648M,2004MNRAS.355..374B, 2004ApJ...612L..93Y,2009ApJ...690.1350O,2012MNRAS.423..862K,2013ApJ...768...71R}. Other possible sources of reionization,
such as quasars and annihilating DM, have little
observational support and would in any case still require, at a minimum, a sizeable
contribution from star-forming galaxies to maintain reionization \citep{1998ApJ...503..505H,2009PhRvD..80c5007B,2010AJ....139..906W,2012MNRAS.425.1413F}. 

There are a variety of constraints on the epoch of
reionization \citep[for a summary of current constraints see][]{2013ApJ...768...71R}; here we will focus
on the observations of the Gunn-Peterson trough \citep{1965ApJ...142.1633G} in quasar spectra \citep{2006AJ....132..117F} and the analysis of the covering fraction of `dark' pixels in quasar spectra \citep{2010MNRAS.407.1328M,2011MNRAS.415.3237M} that
constrains the neutral fraction at the end of reionization, 
plus the Thompson scattering optical
depth of CMB photons that provides an integral constraint over the full
history of reionization. These constraints taken together with the UV-luminosity function argue for an extended period of reionization that
begins early in cosmic time.

Star-forming galaxies during the epoch of reionization must have a significant ionizing-photon escape fraction and the UV galaxy luminosity function must extend beyond the observed limits in both intrinsic luminosity and redshift in order for galaxies to reionize the universe \citep{2012MNRAS.423..862K,2013ApJ...768...71R}. 
The assumed forms of, and values for,  the
parameters used to model reionization -- the redshift evolution of the UV-luminosity function, the limiting
luminosity at which galaxy formation is assumed to truncate, and the
escape fraction of ionizing photons -- have a large impact on the
derived reionization history. Within the CDM paradigm, there exists an
interesting tension between the suppression of star formation in
low-mass DM haloes which is necessary to match near-field observations,
such as the luminosity function and spatial distribution of the Milky Way satellite galaxies, with the expectation of
low-mass galaxies at high redshift to be the dominant source of
reionization \citep{2014MNRAS.443L..44B}. 

As already discussed WDM is a possible solution
to CDM small-scale issues in the local Universe. Recently,
\cite{2014MNRAS.442.1597S} used the predicted high redshift UV-luminosity
functions, reionization history, and CMB optical depth to constrain the mass of a thermally produced WDM particle, ruling out $m_W=1.3\text{ keV}$ at greater than $2 \sigma$ and suggesting sensitivity of future experiments to $m_W= 2.6\text{ keV}$. In this paper, we follow the approach
of \cite{2014MNRAS.442.1597S} in using an abundance-matching technique, albeit 
with a modified procedure that we describe, to predict the high-redshift UV-luminosity
functions, reionization history, and CMB optical depth of axion Mixed
Dark Matter (aMDM) models and compare those predictions to reionization
constraints from observations.

\section{Ultralight axion DM}
\label{sec:Axions} 

We begin this section with a simple argument that relates the relevant mass scales for ULAs and WDM, and then give more details of our semi-analytic model for the ULA mass function.

\subsection{Thermal and non-thermal scales}

Structure formation at late times and on the largest scales
constrains the dominant component of the DM to have growth $\delta\sim a$, where $a$ is the Friedmann--Robertson--Walker scale factor, so that the power spectrum and growth on these largest scales is the same as for CDM. If the DM is not completely cold and pressureless for all of cosmic history, with equation of state $w=P/\rho=0$ and sound speed $c_{\rm s}^2=\delta P/\delta\rho=0$, then scales can be imprinted on structure formation corresponding to the horizon size when any particular change occurred in these quantities.  This scale can be used to suppress the formation of small-scale structure relative to CDM, and thus in hierarchical structure formation suppress the formation of high-redshift galaxies and the onset of reionization.

With WDM \citep[for example a thermal gravitino as in][]{1982PhRvL..48.1636B} the relevant scales are fixed by the temperature, $T$. The equation of state transitions from $w=1/3$ to $w=0$ when the WDM becomes non-relativistic, and structure is suppressed on scales of order the horizon size when $T\sim m_W$. If the DM is non-thermal, as for a ULA or other ultralight scalar, then the relevant scale is the Hubble scale, $H$. The (time averaged) equation of state transitions from $w_a=-1$ to $0$ when the axion mass overcomes Hubble friction in  the Klein--Gordon equation, and structure is suppressed on scales of order the horizon when $H\sim m_a$ \citep[e.g.][]{hu2000,amendola2005,marsh2010}. 

Scales corresponding to dwarf galaxies were horizon size during the radiation dominated era. During this time the temperature is related to the Hubble scale by 
\begin{equation}
T\sim \sqrt{HM_{\rm pl}}\, ,
\label{eqn:t_h_rad_dom}
\end{equation} 
where $M_{\rm pl}=1/\sqrt{8{\rm \pi} G}=2.4\times 10^{18}\text{ GeV}$ is the reduced Planck mass. As already discussed, WDM with (thermal equivalent) mass $m_W\sim \mathcal{O}(1)\text{ keV}$ is a viable candidate to constitute a large fraction of the DM, and may play a role in resolution of the small-scale problems of CDM. If an axion is to affect structure on similar scales, then a simple order of magnitude estimate for the required axion mass is found by relating $T \sim m_W$ and $T\sim (HM_{\rm pl})^{1/2}\sim (m_aM_{\rm pl})^{1/2}$: one finds $m_a\sim 10^{-21} \text{eV}$.

Axions and other light scalar fields with mass in the range $10^{-24}\text{ eV}\lesssim m_a\lesssim 10^{-20}\text{ eV}$ have been called `Fuzzy' (F)CDM \citep{hu2000}. A more detailed study using the full linear transfer function shows that this range of ULA mass affects structure formation on the same scales as WDM with mass in the range $0.1\text{ keV}\lesssim m_W\lesssim 4 \text{ keV}$ \citep{marsh2013b}.

\subsection{The growth of structure and the halo mass function}

Here, we present the Sheth--Tormen \citep{sheth1999} mass function for aMDM including scale-dependent growth: for more details see \cite{marsh2013b}. The transfer functions and growth factor we use are computed using a modified version of \textsc{camb} \citep{camb} which includes light axions and will be described in Marsh et al. (in preparation).

The axion field, $\phi$, evolves according to the Klein--Gordon equation with a quadratic potential,\footnote{This is true near the minimum for the full potential, which is periodic in $\phi$. We ignore anharmonic effects which can be shown to have negligible effect on the Jeans scale for the range of masses we consider \citep{Boyle:2001du}.} $V(\phi)=m_a^2\phi^2/2$. When the field is oscillating about the potential minimum at times $t>t_{\rm osc}$, where $H(t_{\rm osc})\sim m_a$ one can show that the equation of state and sound speed in the effective fluid description \citep{hu1998b} averaged on time periods $t>1/m_a$ are given by \citep[e.g.][]{turner1983b,hu2000,park2012}
\begin{align}
w_a&=0 \, , \\
c_{\rm s}^2&=\frac{k^2/4m_a^2 a^2}{1+k^2/4m_a^2a^2} \, .
\end{align}

\begin{figure*}
\begin{center}
\includegraphics[width=0.49\textwidth]{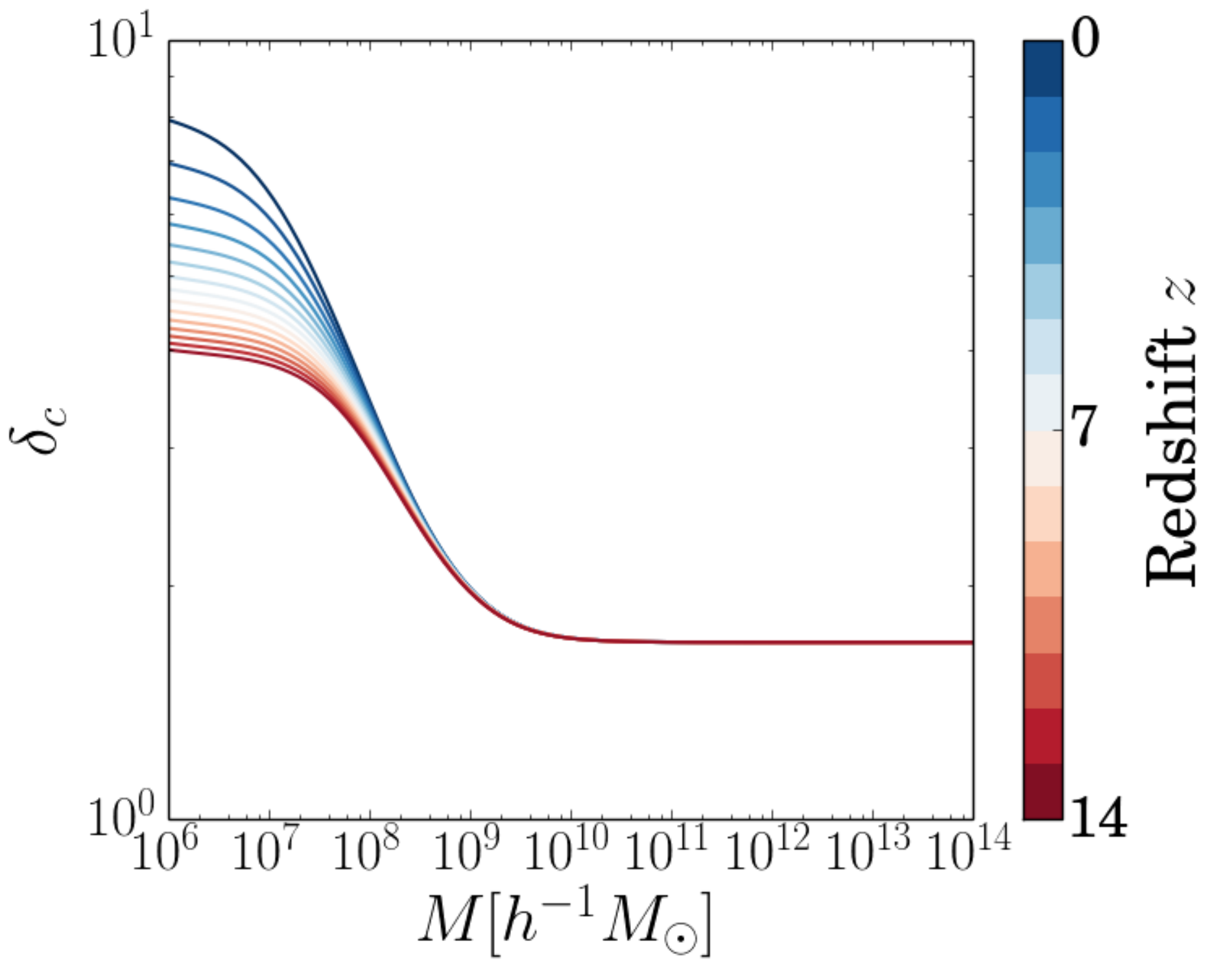}
\includegraphics[width=0.49\textwidth]{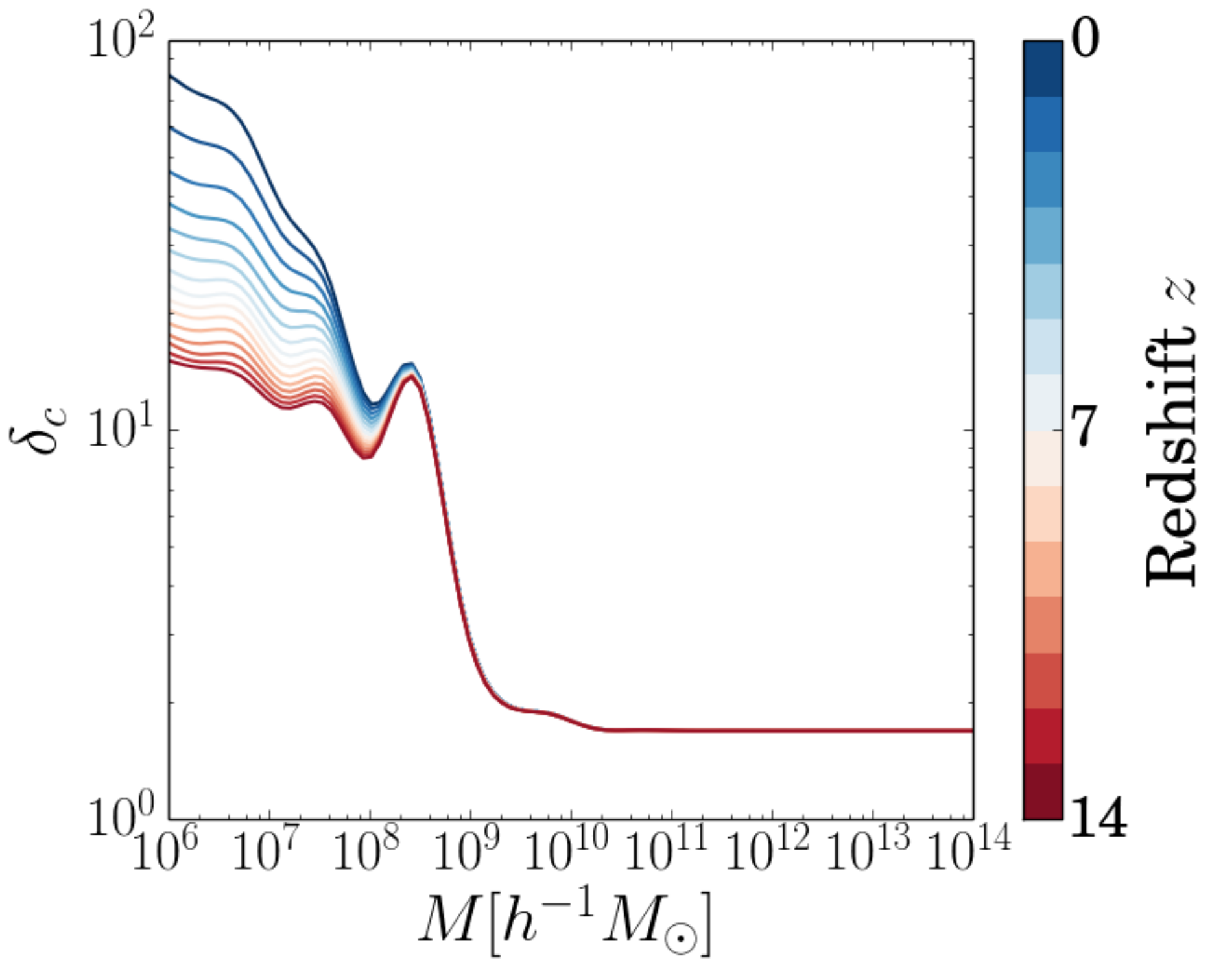} 
\caption{The mass-dependent critical overdensity for two benchmark models in which some or all of the DM is in the form of ULAs, shown for each redshift in the range $0\leq z \leq 14$. Left-hand panel: $m_a=10^{-22}\text{ eV}$, $\Omega_{\rm a}/\Omega_{\rm d}=0.5$. Right-hand panel: $m_a=10^{-22}\text{ eV}$, $\Omega_{\rm a}/\Omega_{\rm d}=1$.}
\label{fig:collapse_barrier}
\end{center}
\end{figure*}

With these prescriptions for the equation of state and sound speed, in a universe dominated by axions at late times, the axion overdensity, $\delta_a=\delta \rho_a/\bar{\rho}_a$, evolves (in the Newtonian gauge and in conformal time) according to \citep{bertschinger1995}
\begin{equation}
\ddot{\delta}_a+\mathcal{H}\dot{\delta}_a + (k^2c_{\rm s}^2-4{\rm \pi} Ga^2\bar{\rho}_a)\delta_a=0 \, ,
\end{equation}
where $\mathcal{H}=aH$ is the conformal Hubble rate and overdots denote derivatives with respect to conformal time, $\eta$. For small $k$, the sound speed goes to zero and we recover scale-independent linear growth, with $\delta_a\sim a$ on large scales. However, for large $k$ the sound speed dominates and the overdensity oscillates rather than grows. The transition between growth and oscillation occurs at the Jeans scale
\begin{equation}
k_{\rm J}=a(16 {\rm \pi} G \bar{\rho}_a)^{1/4}m_a^{1/2} \, .
\end{equation}
For $k>k_{\rm J}$ there is no growth of structure. There is scale-dependent growth as $k$ decreases from $k_{\rm J}$, continuously interpolating to the standard scale-independent linear growth on the largest scales, $k\ll k_{\rm J}$ \citep{1985MNRAS.215..575K}.

In the halo mass function (HMF) one can use the variance of the matter
power spectrum, $\sigma (M)$, computed at redshift $z=0$ if the
barrier for collapse, $\delta_{\rm c}$, is given by the Einstein-de Sitter
value at $z=0$, $\delta_{\rm c,\rm EdS}\approx 1.686$, scaled by the
linear growth: $\delta_{\rm c}(z)=\delta_{\rm c,\rm
EdS}/D(z)$. \cite{marsh2013b} proposed that one could account
for scale-dependent growth by simply replacing $D(z)\rightarrow
D(k,z)$ and then using the enclosed mean mass to define a halo-mass-dependent barrier for collapse, $\delta_{\rm c}(M,z)$. 

Fig.~\ref{fig:collapse_barrier} shows $\delta_{\rm c}(M,z)$ computed in
this manner for two aMDM cosmologies, which can be considered as benchmarks for the purposes of this paper. They each take $m_a=10^{-22}\text{ eV}$ while varying the fractional energy density in axions, $\Omega_{\rm a}=\rho_a/\rho_{\rm crit}$, and CDM, $\Omega_{\rm c}$, and holding the total DM density, $\Omega_{\rm d}=\Omega_{\rm a}+\Omega_{\rm c}$, fixed. The first model takes $\Omega_{\rm a}/\Omega_{\rm d}=0.5$, so that half of the DM is in ULAs, and the second takes $\Omega_{\rm a}/\Omega_{\rm d}=1$. As may be seen in the figure, the barrier for collapse becomes large for low-mass objects due to the vanishing growth on scales below the Jeans scale. This is consistent with what is found from an excursion-set calculation by \cite{2013MNRAS.428.1774B} applied to WDM at the WDM Jeans scale.

\begin{figure*}
\begin{center}
\includegraphics[width=0.49\textwidth]{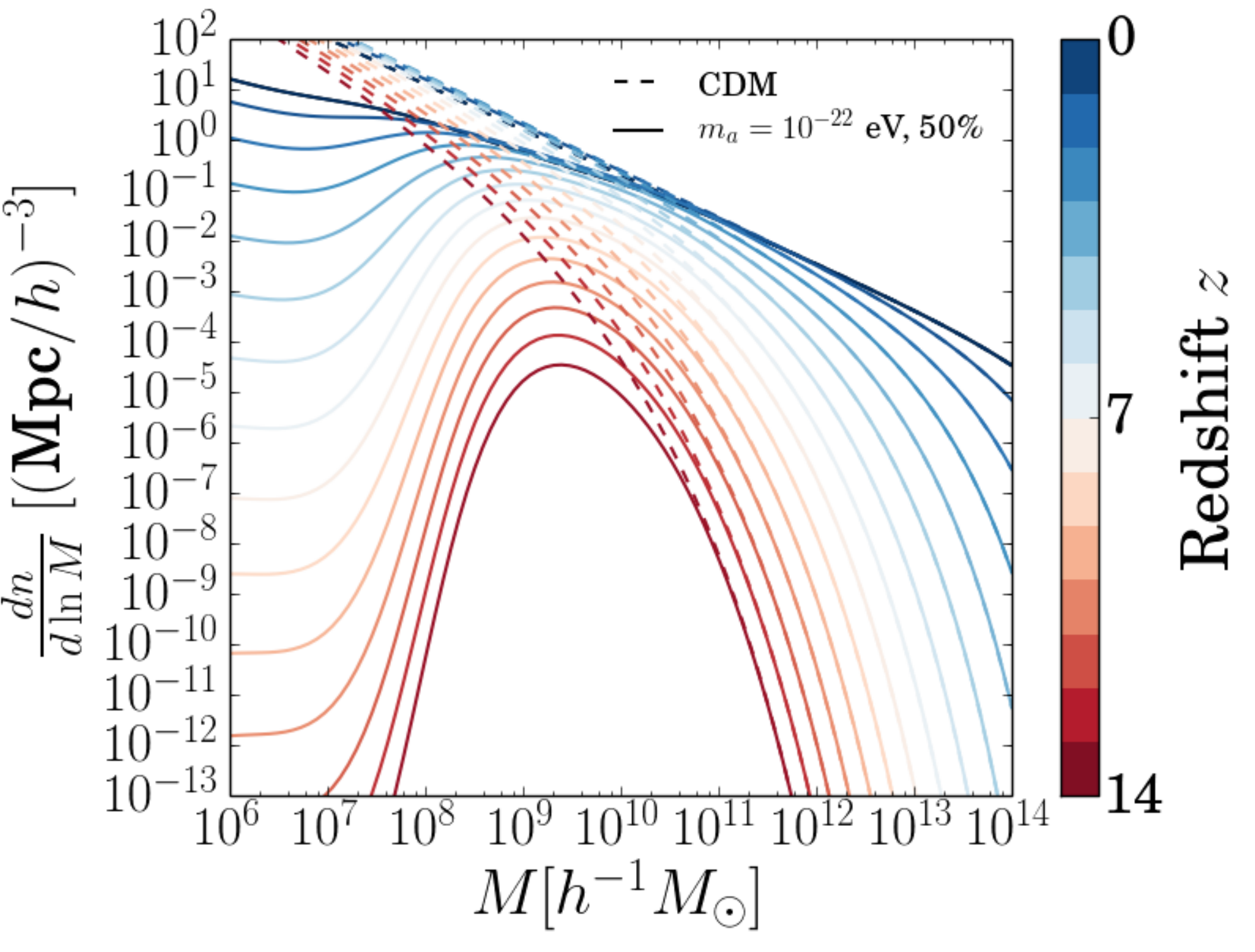}
\includegraphics[width=0.49\textwidth]{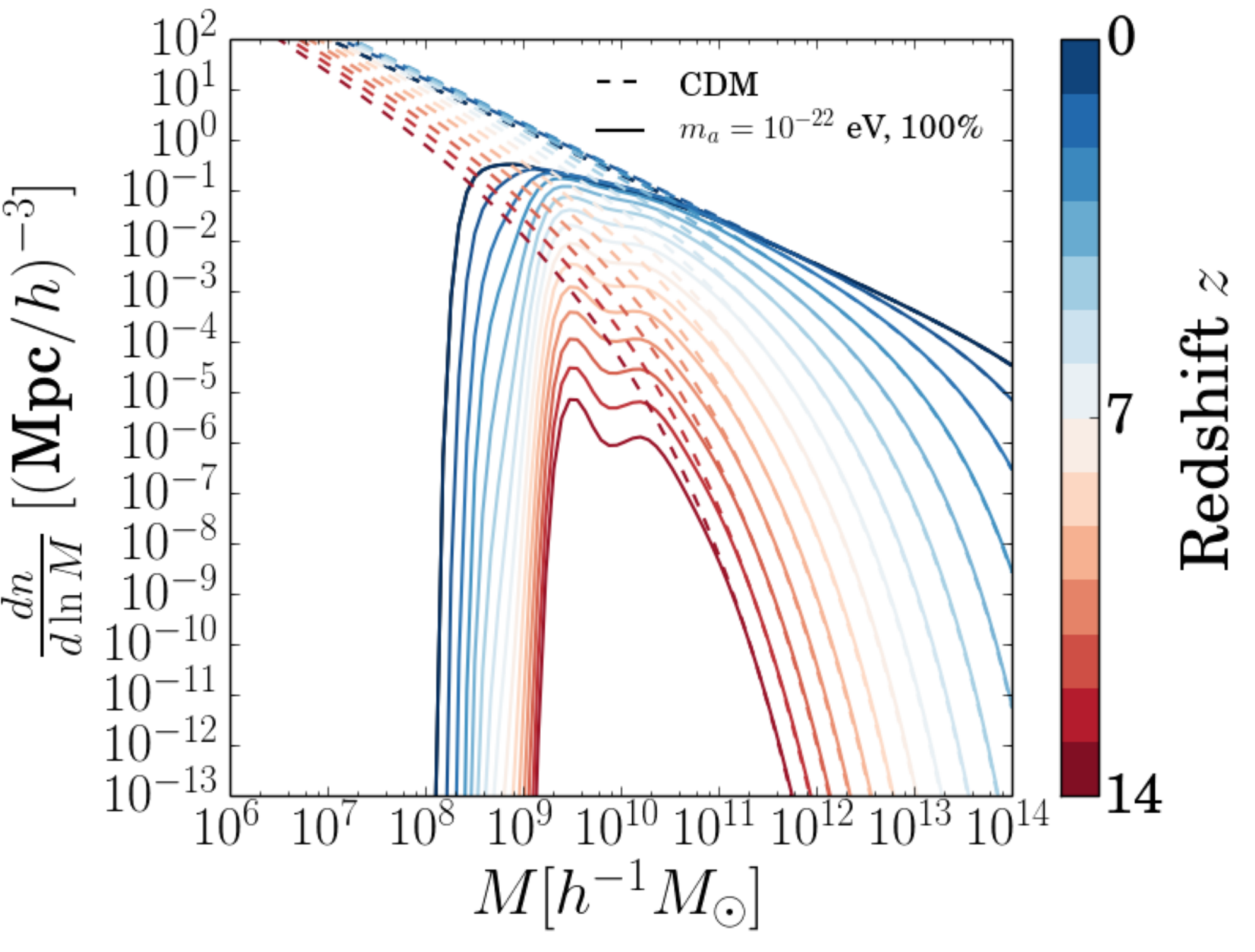} 
\caption{Sheth-Tormen mass function for ULAs including scale-dependent growth, shown for each redshift in the range $0\leq z \leq 14$. The result for CDM is shown for reference. Left-hand panel: $m_a=10^{-22}\text{ eV}$, $\Omega_{\rm a}/\Omega_{\rm d}=0.5$. Right-hand panel: $m_a=10^{-22}\text{ eV}$, $\Omega_{\rm a}/\Omega_{\rm d}=1$.}
\label{fig:hmf_doddy}
\end{center}
\end{figure*}

The mass-dependent barrier for collapse can simply be substituted into
the Sheth--Tormen mass function along with the correct variance to find
${\rm d}n/{\rm d}\ln M$, the number density of haloes per logarithmic mass bin. The HMFs
for the two benchmark cosmologies of
Fig.~\ref{fig:collapse_barrier} are shown in
Fig.~\ref{fig:hmf_doddy}. The rising value of $\delta_{\rm c}(M)$ for low
$M$ is seen to suppress the HMF relative to CDM in both cosmologies,
particularly at high-$z$. The existence of this sharp suppression due
to scale-dependent growth is both a key prediction of aMDM and
the primary means with which we will constrain it. At high redshift, we
should expect many fewer objects to have formed when the DM contains a ULA compared to a pure CDM
universe, even when ULAs are only a fractional component of the DM.

The cut-off in the HMF at $z=13$ in Fig.~\ref{fig:hmf_doddy} (right-hand panel) occurs at $M\approx 10^9 h^{-1}{\rm M}_\odot$, and the HMF peaks near
this value. This is consistent with the high-resolution simulations of
the formation of structure in a universe dominated by axion DM with $m_a=8.1\times
10^{-23}\text{ eV}$ carried out by \cite{schive2014} that report a first
object of mass $M=10^9 M_\odot$ at $z=13$. While \cite{schive2014} do not report the full
HMF from their simulations, we find this quantitative agreement
encouraging as a validation of our semi-analytic model.

\section{UV-Luminosity Functions and the Abundance Matching Technique}
\label{sec:Methods} 

\begin{figure}
\begin{center}
\includegraphics[width=0.49\textwidth]{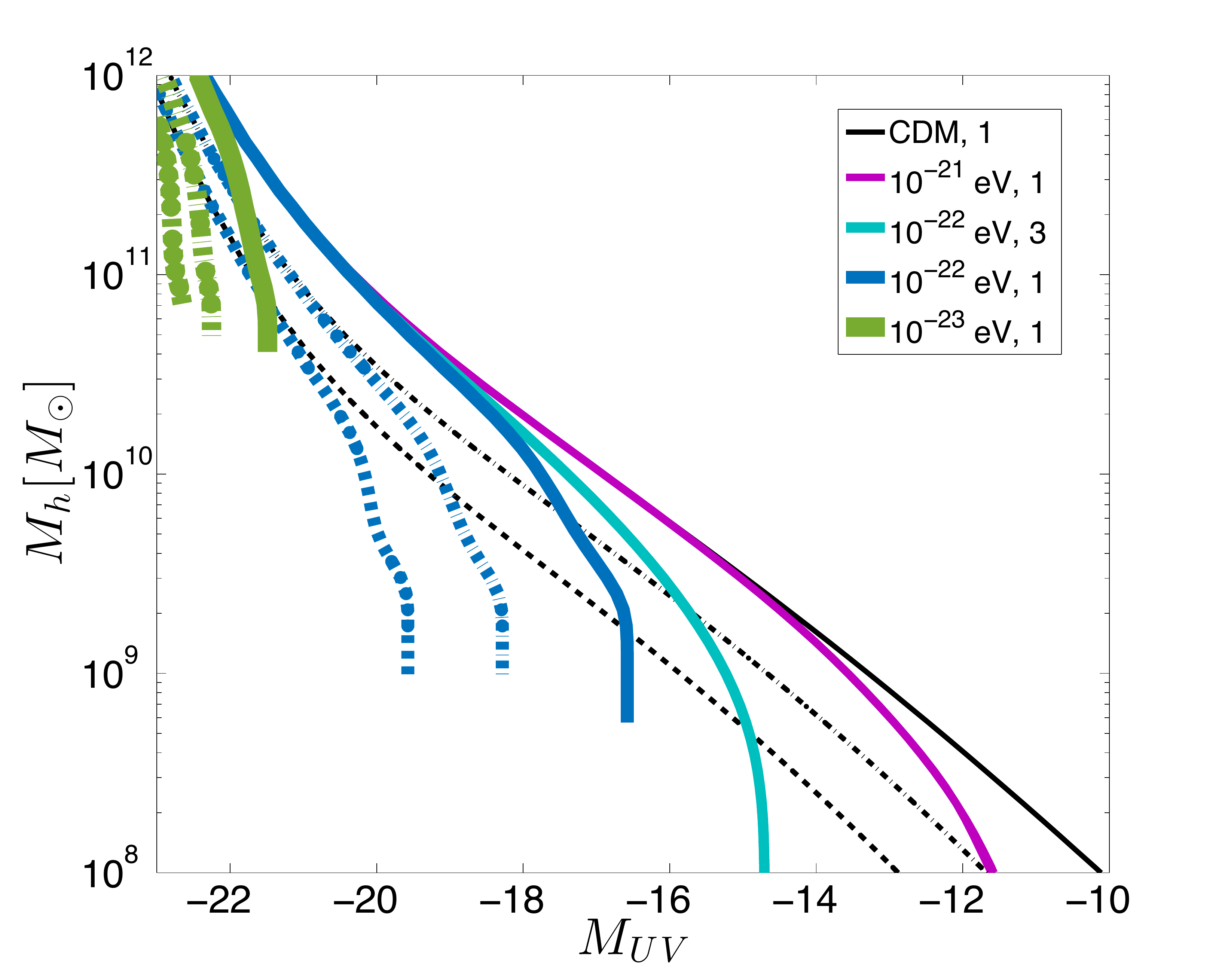} 
\caption{The DM halo mass--galaxy luminosity relation, $M_{\rm h}(M_{\rm UV})$, for CDM (black) and aMDM models $\{$$m_a = 10^{-21} \text{eV}$, `1' (purple); $m_a = 10^{-22} \text{eV}$, `3' (cyan); $m_a = 10^{-22} \text{eV}$, `1' (blue); $m_a = 10^{-23} \text{eV}$, `1' (green)$\}$ at redshifts $z = 7$ (solid curve), $z = 10$ (dot--dashed), and $z=13$ (dashed). The truncation in the $M_{\rm h}(M_{\rm UV})$ relation for models $m_a = 10^{-22} \text{eV}$, `1' and $m_a = 10^{-23} \text{eV}$, `1' is due to a truncation in the corresponding HMF (as shown in the right-hand panel of Fig. \ref{fig:hmf_doddy}). The turnover in the $M_{\rm h}(M_{\rm UV})$ relation in the $m_a = 10^{-22} \text{eV}$, `3' model at $z = 7$ is the result of a turnover (without a complete truncation) in the $m_a = 10^{-22} \text{eV}$, $\Omega_{\rm a}/\Omega_{\rm d}=0.5$ HMF at $z = 7$ (left-hand panel of Fig. \ref{fig:hmf_doddy}).}
\label{fig:Mh_M}
\end{center}
\end{figure}

We use the abundance-matching technique \citep{2004ApJ...609...35K,2004MNRAS.353..189V}
to connect the aMDM HMFs discussed in Section~\ref{sec:Axions} to high-redshift
rest-frame UV-luminosity functions determined from deep imaging in optical to near-IR bandpasses of \textit{HST} Legacy Fields (\citealt{2014arXiv1403.4295B} and references therein). The abundance-matching technique assigns a galaxy of a given absolute rest-frame UV (1500 {\AA}) magnitude, $M_{\rm UV}$ (we will use AB-magnitudes for the purposes of comparisons with observational data), to a given DM halo mass, $M_{\rm h}$, by assuming that each DM halo hosts one galaxy and that the relationship between dark-matter halo
mass and galaxy luminosity, $M_{\rm h}(M_{\rm UV})$, is monotonic.

The first step in this matching process is characterization of the galaxy luminosity function (the number
density of galaxies per absolute magnitude), $\phi(M)$, by fitting a suitable analytic function to the \textit{HST} galaxy number counts at redshifts
$z=6$--$10$. We adopt the usual practice of fitting a Schechter function \citep{1976ApJ...203..297S} which  has the
following form:
\begin{equation}
\phi(M) = \phi_{\star}(\frac{\ln(10)}{2.5})10^{-0.4(M-M_{\star})\alpha}\exp(-10^{-0.4(M-M_{\star})}),
\label{eqn:schecht}
\end{equation}

\noindent where $\phi_{\star}$ is the normalization, $M_{\star}$ is
the characteristic magnitude, $\alpha$ is the faint-end slope, and $M_{\rm UV}$ is used when applying equation (\ref{eqn:schecht}). We use two sets of Schechter function parameters, respectively, taken from
\cite{2014arXiv1403.4295B} and \cite{2012MNRAS.423..862K} (their `FIT'
model), since different values for the Schechter function parameters 
(particularly the faint-end slope) can have a significant effect on
the resulting reionization history. Each Schechter function fit is
extrapolated to fainter magnitudes and redshifts where there are not
currently observations by assuming the values of the parameters evolve linearly with
redshift consistent with the trends in the data at redshifts 6--10 (see the above cited works for the model details).  
The data that the \cite{2014arXiv1403.4295B} luminosity function is based on
includes more recent data than that of \cite{2012MNRAS.423..862K}, but both models are consistent with the current data set.

The parametrized fit to the observed galaxy luminosity
function and the DM HMF of a given model  are, at
each redshift, integrated to obtain, respectively, the
cumulative galaxy luminosity function,
$\Phi(<M_{\rm UV})$, the number density of galaxies brighter than $M_{\rm UV}$
and the cumulative DM HMF, $n(>M_{\rm h})$, the number density of haloes more massive than $M_{\rm h}$. For each DM model, an absolute magnitude, $M_{\rm UV}$, is assigned to a dark
matter halo mass, $M_{\rm h}$ by matching number densities in the cumulative functions, i.e. according to the relation:
\begin{equation}
\Phi(<M_{\rm UV},z) = n(>M_{\rm h},z) .
\end{equation} 
This  gives the DM halo mass--galaxy luminosity relations,
$M_{\rm h}(M_{\rm UV})$, shown in Figure \ref{fig:Mh_M}. The
$M_{\rm h}(M_{\rm UV})$ relation is then used to convert the cumulative DM mass function of a given model into a predicted cumulative galaxy luminosity function. 

This may appear to be a circular process but the predicted cumulative luminosity function for each DM model will match exactly with the input cumulative galaxy luminosity function
derived from observations only provided that the DM HMF actually contain low-mass haloes of a sufficient (cumulative) number density to match the faint end of the observed luminosity function -- otherwise the predicted luminosity function will end prematurely compared to observations. 

Indeed, a truncation in the HMF at some minimum halo mass, as shown in the right-hand panel of Fig.~\ref{fig:hmf_doddy}, leads to a corresponding truncation in the $M_{\rm h}(M_{\rm UV})$ relation, as is clearly seen for the $m_a=10^{-22}\text{
eV}$, Model 1 (100 per cent axion DM), case in Fig. \ref{fig:Mh_M}. For the case of a turnover in the HMF without a complete truncation, as shown in the left-hand panel of Fig.~\ref{fig:hmf_doddy}, the $M_{\rm h}(M_{\rm UV})$ relation will steepen such that several orders of magnitude in DM halo mass maps on to a nearly singular value of galaxy luminosity, as can be seen for the $m_a=10^{-22}\text{eV}$, Model 3 (50 per cent axion DM), case in Fig. \ref{fig:Mh_M}. A truncation will occur in
the resulting aMDM cumulative luminosity function at the corresponding
magnitude for both cases. The terminal value in the aMDM cumulative luminosity function, therefore, indicates the minimum mass scale of galaxy formation at each redshift based on whether a sufficient number of DM haloes of that mass scale have
collapsed. 

The advantage of the abundance-matching procedure is that it provides a
pathway to constraining DM mass functions by directly
comparing to galaxy observations without appealing to uncertain galaxy
formation physics. The $M_{\rm h}(M_{\rm UV})$
relation additionally serves as a prediction for validation or
rejection of a given theory. 

\cite{2014MNRAS.442.1597S} used a
different methodology in their abundance-matching procedure for the
WDM case. Those authors used the $M_{\rm h}(M_{\rm UV})$ relation obtained from
the CDM abundance-matching when constructing the predicted WDM
cumulative luminosity functions. Their argument for this choice was
the unknown galaxy formation physics that accounts for their
$M_{\rm h}(M_{\rm UV})$ relation should be based on CDM, as WDM mass functions
would require a more efficient galaxy formation process in low-mass
galaxies. Our approach uses the same DM mass function at the
beginning and end of the abundance-matching procedure, which we 
consider to be more self-consistent. 

\section{Reionization}
\label{sec:Methods II} 
 
We determine the reionization history of aMDM models, as represented by the volume-filling fraction of ionized hydrogen, $Q_{\text{H\rom{2}}}(z)$. The volume-filling fraction of ionized hydrogen balances the ionization of the neutral intergalactic medium (IGM) with the recombination of free electrons and protons, as given by the differential equation \citep{1999ApJ...514..648M,2012MNRAS.423..862K, 2013ApJ...768...71R,2014MNRAS.442.1597S}: 
\begin{equation}
\frac{\mathrm{d}Q_{\text{H\rom{2}}}}{\mathrm{d}t} = \frac{\dot{n}_{\rm ion}}{\overline{n}_{\rm H}} - \frac{Q_{\text{H\rom{2}}}}{\overline{t}_{\rm rec}},
\end{equation} 
where the $\overline{n}_{\rm H}$ term represents the mean comoving hydrogen number density 
and $\dot{n}_{\rm ion}$ is the comoving production rate of ionizing photons per unit volume. The parameter $\overline{t}_{\rm rec}$ is the volume-averaged recombination time of ionized hydrogen given by the equation:
\begin{equation}
\overline{t}_{\rm rec} = \frac{1}{C_{\text{H\rom{2}}}\alpha_{B}(T_{0})\overline{n}_{\rm H}(1+Y/4X)(1+z)^3},
\end{equation}
where $C_{\text{H\rom{2}}} \equiv \frac{<n_{\rm H}^2>}{<n_{\rm H}>^2}$ is the clumping factor of ionized gas,
$\alpha_B(T_0)$ is the case B hydrogen recombination coefficient for
an IGM temperature of $T_0$, and X and $Y = 1-X$ are, respectively, the
primordial hydrogen and helium abundances. The appropriate value of the clumping
factor of ionized gas is uncertain and varies based on definition and
method \citep[see][and references therein]{2013ApJ...768...71R}. We therefore follow the literature and
choose a value of $C_{\text{H\rom{2}}} = 3$ \citep{2012MNRAS.423..862K,2014MNRAS.442.1597S}. We also follow previous work in adopting the commonly assumed values of $T_0 = 2\times10^4 \text{ K}$, $X = 0.76$ and $Y = 0.24$ \citep{2012MNRAS.423..862K,2014MNRAS.442.1597S}. The primordial helium and hydrogen abundances are consistent with both CMB measurements \citep{2011ApJS..192...18K} and estimates from low-metallicity extragalactic regions \citep{2004ApJ...602..200I,2007ARNPS..57..463S}. The assumed IGM temperature is appropriate for ionized gas at the mean density during the epoch of reionization \citep{2003ApJ...596....9H}.

\begin{figure*}
\begin{center}
\includegraphics[width=0.49\textwidth]{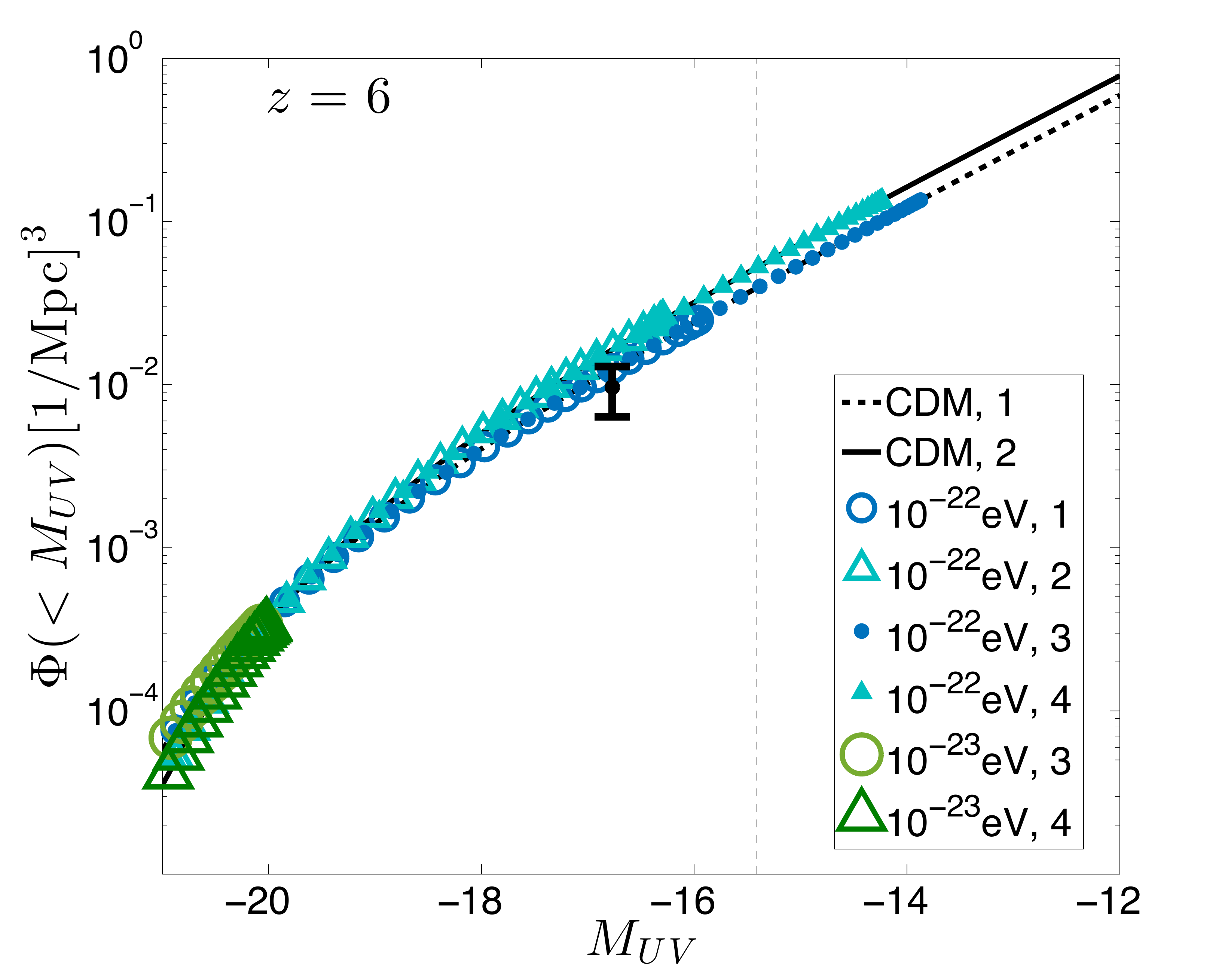} 
\includegraphics[width=0.49\textwidth]{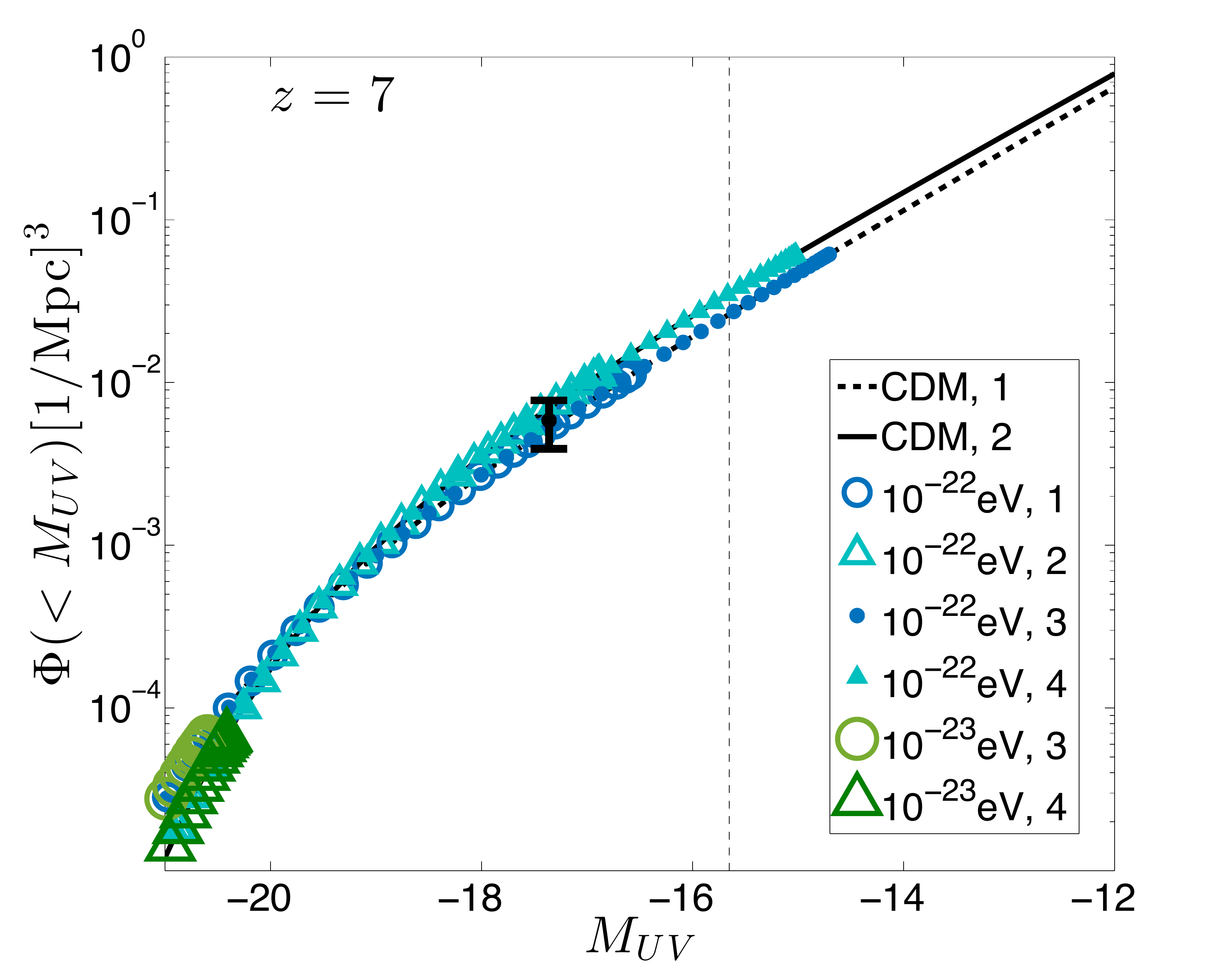} 
\includegraphics[width=0.49\textwidth]{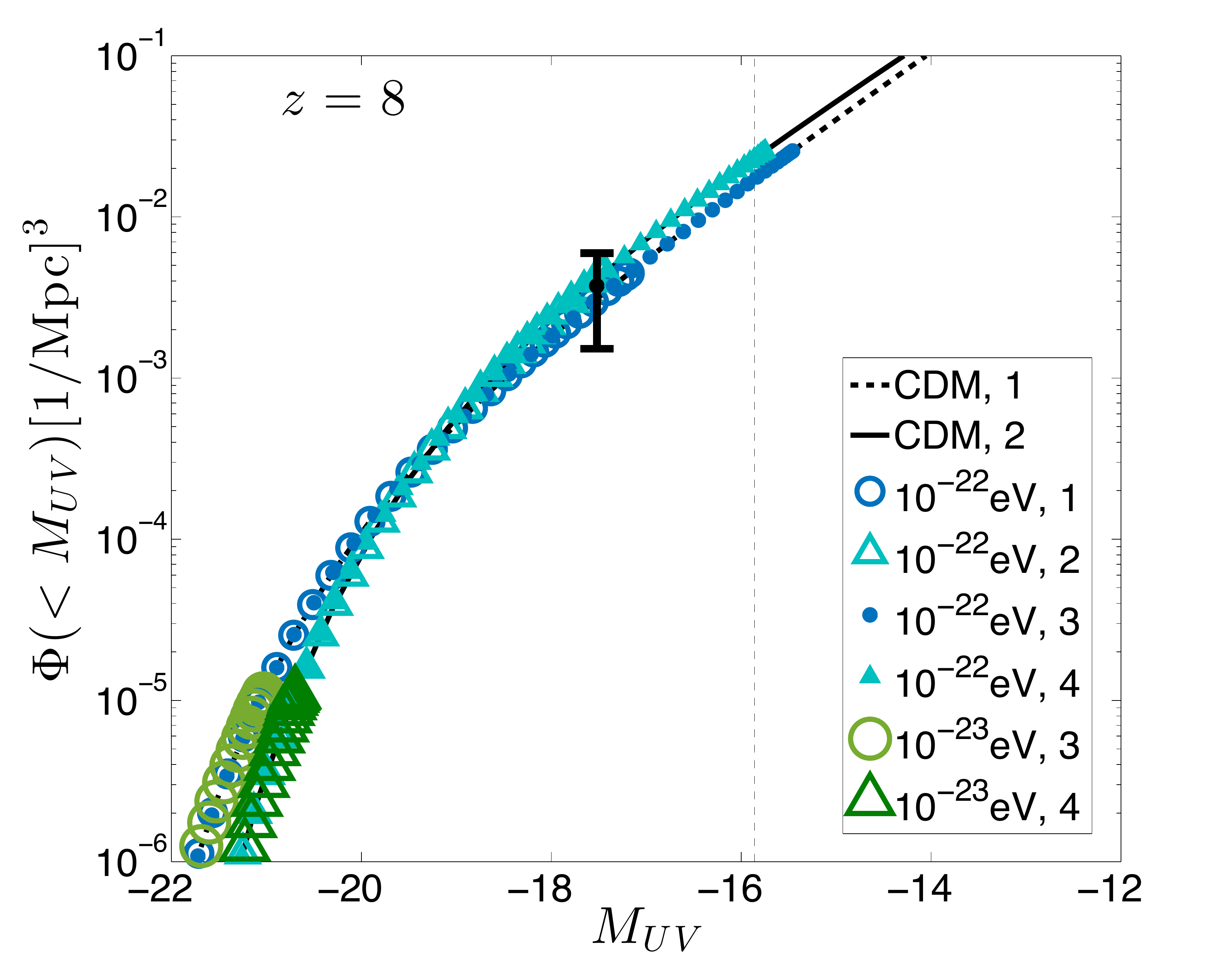} 
\includegraphics[width=0.49\textwidth]{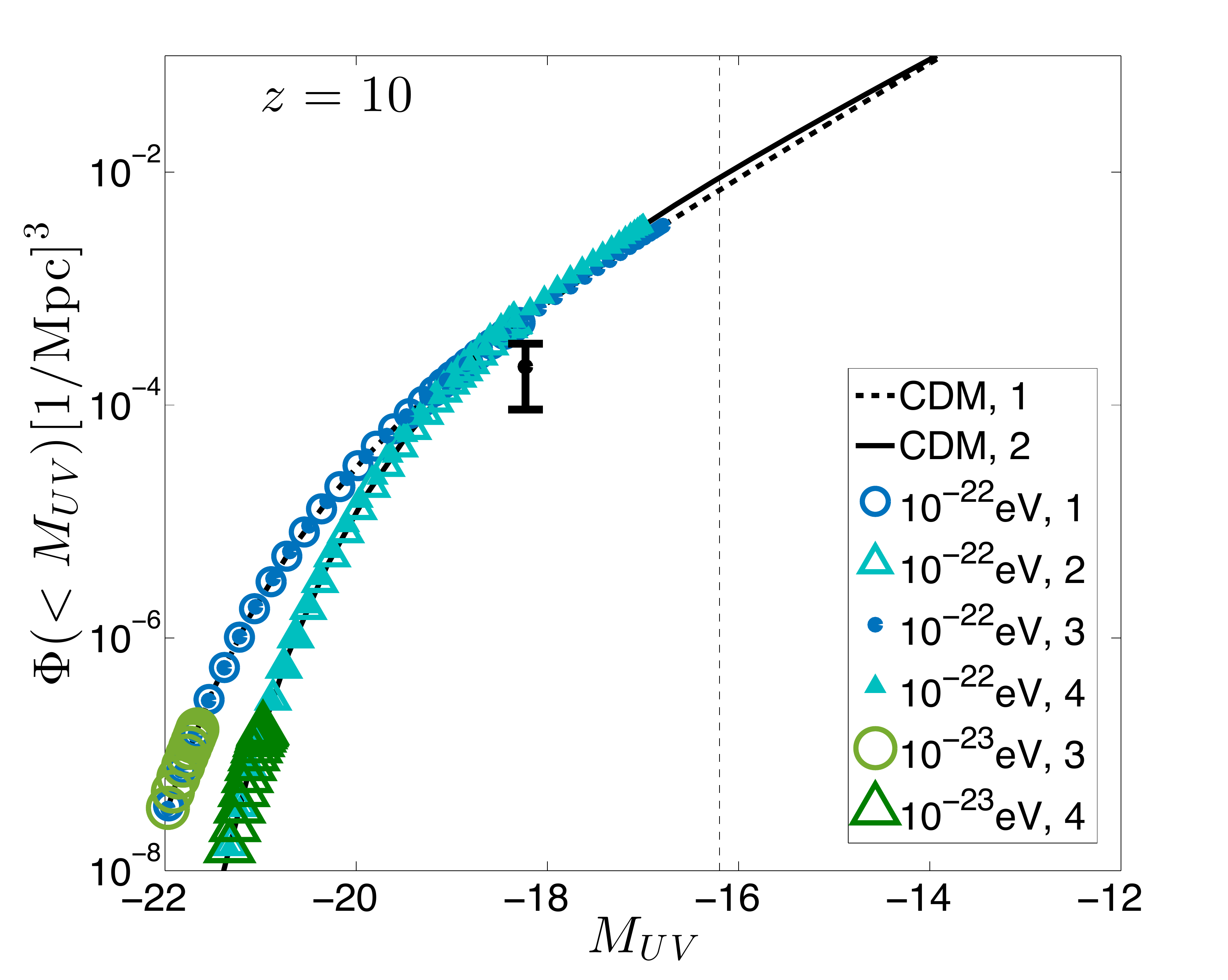} 
\caption{The cumulative luminosity functions ($z = 6, 7, 8, \text{and }10$) of CDM models `1' (dashed black) and `2' (solid black) and aMDM $m_a=10^{-23}\text{ eV}$ models `3' (large green circle) and `4' (large green triangle), and aMDM $m_a=10^{-22}\text{ eV}$ models `1' (medium blue circle), `2' (medium blue triangle), `3' (small filled blue circle), `4' (small filled blue triangle). The data points on each plot are the cumulative number density of galaxies in \textit{HST} fields \citep{2014arXiv1403.4295B} summed down to the faint-end limit at each redshift. The error bars are $2\sigma$ for $z$ = 6, 7, and 8 ($1\sigma$ for $z$ = 10). The dashed vertical line in each panel is the absolute magnitude faint-end limit \textit{JWST} will reach at each redshift for a survey down to an apparent magnitude of $\text{AB} = 31.5\text{ mag}$ \citep{2006NewAR..50..113W}. The $m_a=10^{-23}\text{ eV}$ aMDM models truncate prior to reaching the HUDF faint-end limit, and thus this model is ruled out ($>8\sigma$). The truncation magnitude for the $m_a=10^{-22}\text{ eV}$ models scales according to $\Omega_{\rm a}/\Omega_{\rm d}$, but all axion fractions of DM are consistent with HUDF constraints. The $m_a=10^{-21}\text{ eV}$ models are not shown on this plot as they are indistinguishable from CDM over the scales shown. Note: The \textit{x}-axis and \textit{y}-axis limits are different in each panel.}
\label{fig:Lum_z6_8}
\end{center}
\end{figure*}

The production rate of ionizing photons, $\dot{n}_{\rm ion}$, is given by the equation:
\begin{equation}
\dot{n}_{\rm ion} = f_{\rm esc}\int_{M_{\rm lim}}^\infty  \! \phi(M_{\rm UV})\gamma_{\rm ion}(M_{\rm UV}) \, \mathrm{d}M_{\rm UV},
 \label{eqn:nion}
\end{equation} 
where $\phi(M_{\rm UV})$ is the galaxy UV-luminosity function given in equation (\ref{eqn:schecht}), $\gamma_{\rm ion}(M_{\rm UV})$ is a conversion factor that converts the galactic UV-luminosity to hydrogen ionizing photon luminosity, and $f_{\rm esc}$ represents the escape fraction of ionizing photons. We use equation 6 of \citet{2012MNRAS.423..862K} that defines:
\begin{equation}
\gamma_{\rm ion}(M_{\rm UV}) \equiv 2\times10^{25}\text{ s}^{-1} (10^{0.4(51.63-M_{\rm UV})}) \zeta_{\rm ion}, 
\end{equation}
where the second term is the galactic rest-frame UV (1500 {\AA}) luminosity using AB magnitudes, and $\zeta_{\rm ion}$ is a dimensionless parameter that contains all assumptions of the galaxy stellar spectrum properties, e.g. the slope of the UV continuum. We adopt the fiducial model of \citet{2012MNRAS.423..862K} and set $\zeta_{\rm ion} = 1$. The values of $M_{\rm lim}$ and $f_{\rm esc}$ are model parameters that we allow to vary in our analysis. For aMDM models where the $M_{\rm h}(M_{\rm UV})$ relation truncates prior to reaching $M_{\rm lim}$, equation (\ref{eqn:nion}) is integrated down to the truncation magnitude in place of $M_{\rm lim}$.

The most robust  constraint on the epoch of reionization is the CMB Thompson scattering optical depth, $\tau$. The CMB optical depth is an integral over the reionization history to redshift $z$, given by the equation:
\begin{equation}
\tau(z) = \int_{0}^{z}  \! \frac{c(1+z')^2}{H(z')}Q_{\text{H\rom{2}}}(z')\sigma_{\rm T}\overline{n}_{\rm H}(1+\eta Y/4X) \, \mathrm{d}z',§	
\end{equation}
where $c$ is the speed of light, $H(z)$ is the Hubble parameter, $\sigma_{\rm T}$ is the Thompson scattering cross-section, and $\eta$ gives the ionization state of helium. Following \citet{2012MNRAS.423..862K}, we assume helium is singly ionized ($\eta = 1$ ) at $z>4$ and doubly ionized ($\eta = 2$) at $z\leq4$. 

\section{Results}
\label{sec:Results} 

\subsection{aMDM cumulative luminosity functions}
\label{sec:cum_lum}

\begin{figure}
\begin{center}
\includegraphics[width=0.49\textwidth]{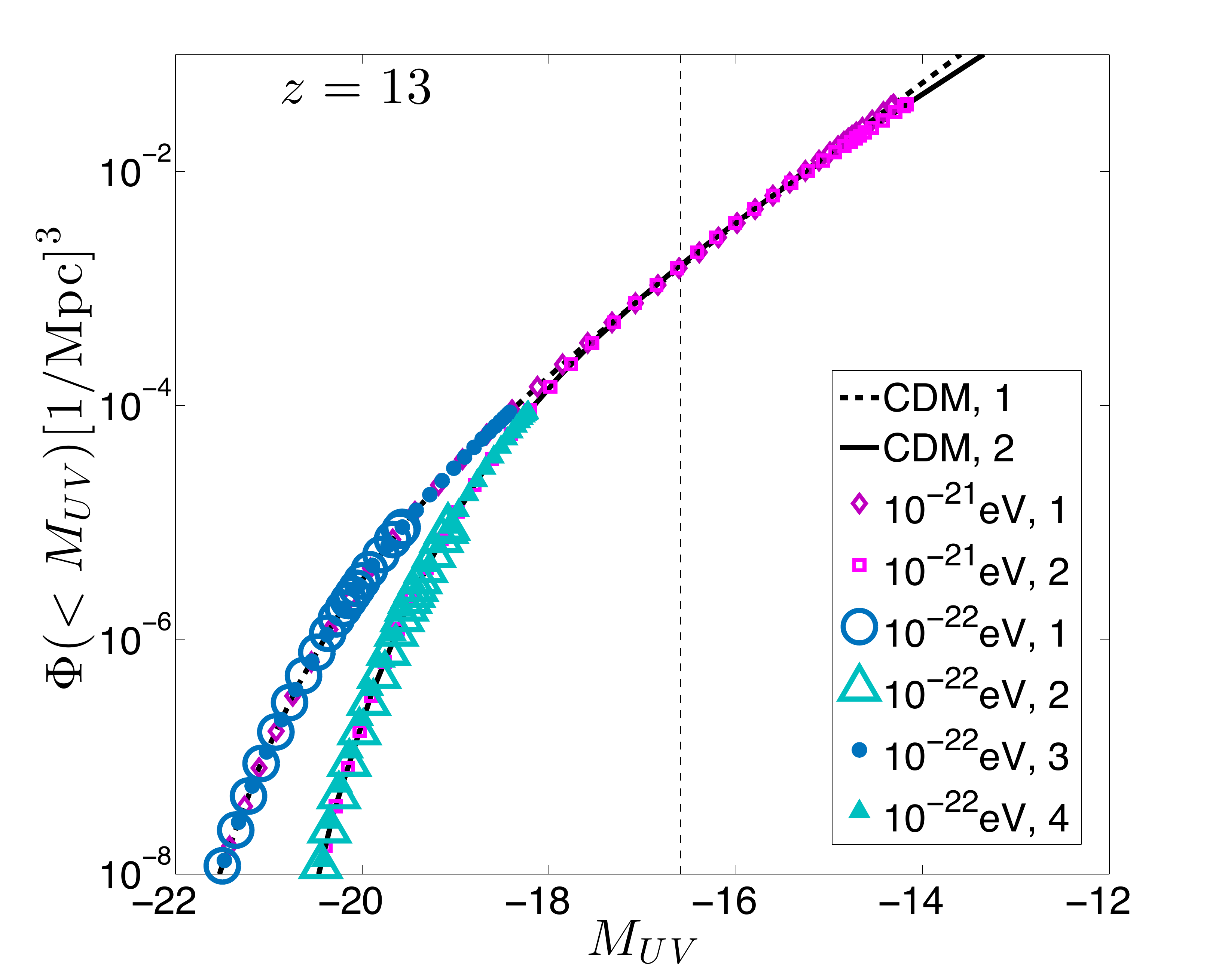} 
\caption{The cumulative luminosity functions ($z = 13$) of CDM models `1' (dashed black) and `2' (solid black) and aMDM $m_a=10^{-21}\text{ eV}$ models `1' (small purple diamond) and `2' (small purple square), and aMDM $m_a=10^{-22}\text{ eV}$ models `1' (medium blue circle), `2' (medium blue triangle), `3' (small filled blue circle), `4' (small filled blue triangle). The $m_a=10^{-23}\text{ eV}$ models are not shown as their maximum cumulative luminosity value fall below the \textit{y}-axis minimum. The dashed vertical line is the absolute magnitude faint-end limit \textit{JWST} will reach for a survey down to an apparent magnitude of $\text{AB} = 31.5\text{ mag}$ \citep{2006NewAR..50..113W}. \textit{JWST} observations with this sensitivity will be able to constrain $m_a=10^{-22}\text{ eV}$ models, but will be unable to distinguish $m_a=10^{-21}\text{ eV}$ models from CDM.}
\label{fig:Lum_z10_13}
\end{center}
\end{figure}

The aMDM cumulative luminosity functions of the $m_a=10^{-21}\text{ eV},10^{-22}\text{ eV}, \text{and }10^{-23}\text{ eV}$ models for the redshifts $z = 6,7,8,10,\text{and }13$ are shown in Figs~\ref{fig:Lum_z6_8} and \ref{fig:Lum_z10_13}. For each axion mass model, we allow four model parameters to vary: the Schechter function fit, the axion fraction of DM ($\Omega_{\rm a}/\Omega_{\rm d}$), the escape fraction of ionizing photons ($f_{\rm esc}$), and the minimum UV magnitude ($M_{\rm lim}$). Table \ref{tab:modellist} lists the model label and set of parameter values used in each model. The model number (`1' to `4') refers to the Schechter function fit and the axion fraction of DM, while the letter in the model label (`a--d') refers to the model reionization parameters: $f_{\rm esc}$ and $M_{\rm lim}$. We will ignore the letter in the model label in this section as the two model parameters it references do not affect the cumulative luminosity functions for the range of magnitudes shown.

The CDM cumulative luminosity functions at each redshift are also plotted in Figs.~\ref{fig:Lum_z6_8} and  \ref{fig:Lum_z10_13}. The two Schechter function parameterizations used in our analysis produce similar CDM cumulative luminosity functions (CDM models `1' and `2') below $z\leq8$, but differ at the bright end for $z>8$ (as seen in Fig. \ref{fig:Lum_z10_13}). We explore how these differences in the UV-luminosity function can produce large differences in the reionization history in the next section.

The cumulative luminosity functions for all $m_a=10^{-23}\text{ eV}$ models truncate prior to reaching the Hubble Ultra Deep Field (HUDF) faint-end limit at all redshifts greater than $z=6$. When this axion is 100 per cent of the DM the cumulative luminosity functions truncate where there are no low mass haloes to host galaxies with fainter magnitudes in the redshift range $z=6$--$13$ (models 1 and 2 therefore do not appear in the plots), while some haloes are produced when this axion is 50 per cent of the DM (models 3 and 4). The DM HMFs of models `3' and `4' at $z= 6$ are suppressed at the low-mass end such that, while there are low-mass haloes below the mass associated with the cumulative luminosity function truncation magnitude, there is not a sufficient number of haloes to host the inferred number of faint galaxies. 

The truncation magnitude will vary somewhat depending on the Schechter function model, but in the models we consider the variation is small and is unable to bring $m_a=10^{-23}\text{ eV}$ into agreement with the data. The cumulative luminosity function (\textit{y}-axis) value at truncation in each panel of Fig.~\ref{fig:Lum_z6_8} gives the total abundance of galaxies at that redshift and must reach the HUDF data point in order to account for the currently observed number count of galaxies. Falling below the data point indicates the model predicts fewer galaxies than are already observed. We use the error bars in Fig.~\ref{fig:Lum_z6_8} to calculate the $\chi^2$(following the method of \citealt{2014MNRAS.442.1597S}) to quantify disagreement of the $m_a=10^{-23}\text{ eV}$ aMDM models with HUDF data. The HUDF data rules out ULAs with $m_a\leq 10^{-23}\text{ eV}$ from contributing more than half of the total DM at greater than $8\sigma$.

\begin{table}
\begin{center}
\caption{Axion mixed DM models}  \label{tab:modellist}
\begin{tabular}{l | l | r | r | r }
Model & Schechter fit & $\Omega_{\rm a}/\Omega_{\rm d}$ & $f_{\rm esc}$ & $M_{\rm lim}$ \\ \hline
1a & Bouwens & 1.0 & 0.2 & -13  \\  \hline
1b & Bouwens & 1.0 & 0.2 & -10  \\  \hline
1c & Bouwens & 1.0 & 0.5 & -13  \\  \hline
1d & Bouwens & 1.0 & 0.5 & -10  \\  \hline
2a & Kuhlen & 1.0 & 0.2 & -13  \\  \hline
2b & Kuhlen & 1.0 & 0.2 & -10  \\  \hline
2c & Kuhlen & 1.0 & 0.5 & -13  \\  \hline
2d & Kuhlen & 1.0 & 0.5 & -10  \\  \hline
3a & Bouwens & 0.5 & 0.2 & -13  \\  \hline
3b & Bouwens & 0.5 & 0.2 & -10  \\  \hline
3c & Bouwens & 0.5 & 0.5 & -13  \\  \hline
3d & Bouwens & 0.5 & 0.5 & -10  \\  \hline
4a & Kuhlen & 0.5 & 0.2 & -13  \\  \hline
4b & Kuhlen & 0.5 & 0.2 & -10  \\  \hline
4c & Kuhlen & 0.5 & 0.5 & -13  \\  \hline
4d & Kuhlen & 0.5 & 0.5 & -10  \\  \hline
\end{tabular}
\\
\end{center}
Column 1: Model label, Column 2: Schechter function parameter set, Column 3: axion fraction of DM, Column 4: escape fraction of ionizing photons, Column 5: limiting magnitude of UV-luminosity function. The alphabetic order of the letters in column 1 signifies a progression in the reionization parameter assumptions from most conservative to least conservative, such that `a' indicates the most conservative reionization assumptions (i.e. the smallest escape fraction and brightest limiting magnitude) and `d' corresponds to the least conservative assumptions. CDM models use only the 1a--1d and 2a--2d labels.
\end{table}

The $m_a=10^{-22}\text{ eV}$ and $m_a=10^{-21}\text{ eV}$ aMDM model cumulative luminosity functions, for all axion fractions of DM, are indistinguishable from CDM down to magnitudes fainter than the HUDF faint-end limit for the redshifts $z=6$--$10$. The $m_a=10^{-21}\text{ eV}$ aMDM models are not shown for $z=6$--$10$ as they are consistent with CDM for all magnitudes plotted at these redshifts. We show the cumulative luminosity functions of $m_a=10^{-21}\text{ eV}$ models `1' and  `2'  at $z=13$ in Fig. \ref{fig:Lum_z10_13}. The cumulative luminosity functions truncate at a magnitude of $M_{\rm UV}\approx-14$ distinguishing this mass from CDM at high-$z$. 

The $m_a=10^{-22} \text{ eV}$ cumulative luminosity functions for models `1' and `2' (where ULAs account for all of the DM) truncate at a magnitude only slightly fainter than the HUDF limit for $z=8$ and $10$. The dashed vertical lines in Fig. \ref{fig:Lum_z6_8} show the faint-end limit of a \textit{JWST} `deep field' survey for the redshift range $z=6$--$13$ down to an apparent magnitude of AB = 31.5 mag. At this limiting magnitude it is not possible to distinguish CDM from $m_a=10^{-21}\text{ eV}$ at $z = 13$. The ULA model with $m_a=10^{-22}$, however, predicts that no galaxies with limiting magnitude $M_{\rm UV}\approx -16$ should be seen at high-$z$, and therefore a non-observation by \textit{JWST} would provide evidence that the DM could be composed of such a ULA. A non-observation by JWST at $M_{\rm UV}\approx -16$ does not rule out CDM or $m_a=10^{-21}$ models, but would require a physical explanation for suppression of galaxy formation at that magnitude. On the other hand, if \textit{JWST} does observe galaxies with limiting magnitude $M_{\rm UV}\approx -16$ at high-$z$ this would rule out $m_a=10^{-22}\text{ eV}$ as a dominant component of the DM.

\subsection{aMDM reionization history}
\label{sec:q_hist}

\begin{figure*}
\begin{center}
\includegraphics[width=0.49\textwidth]{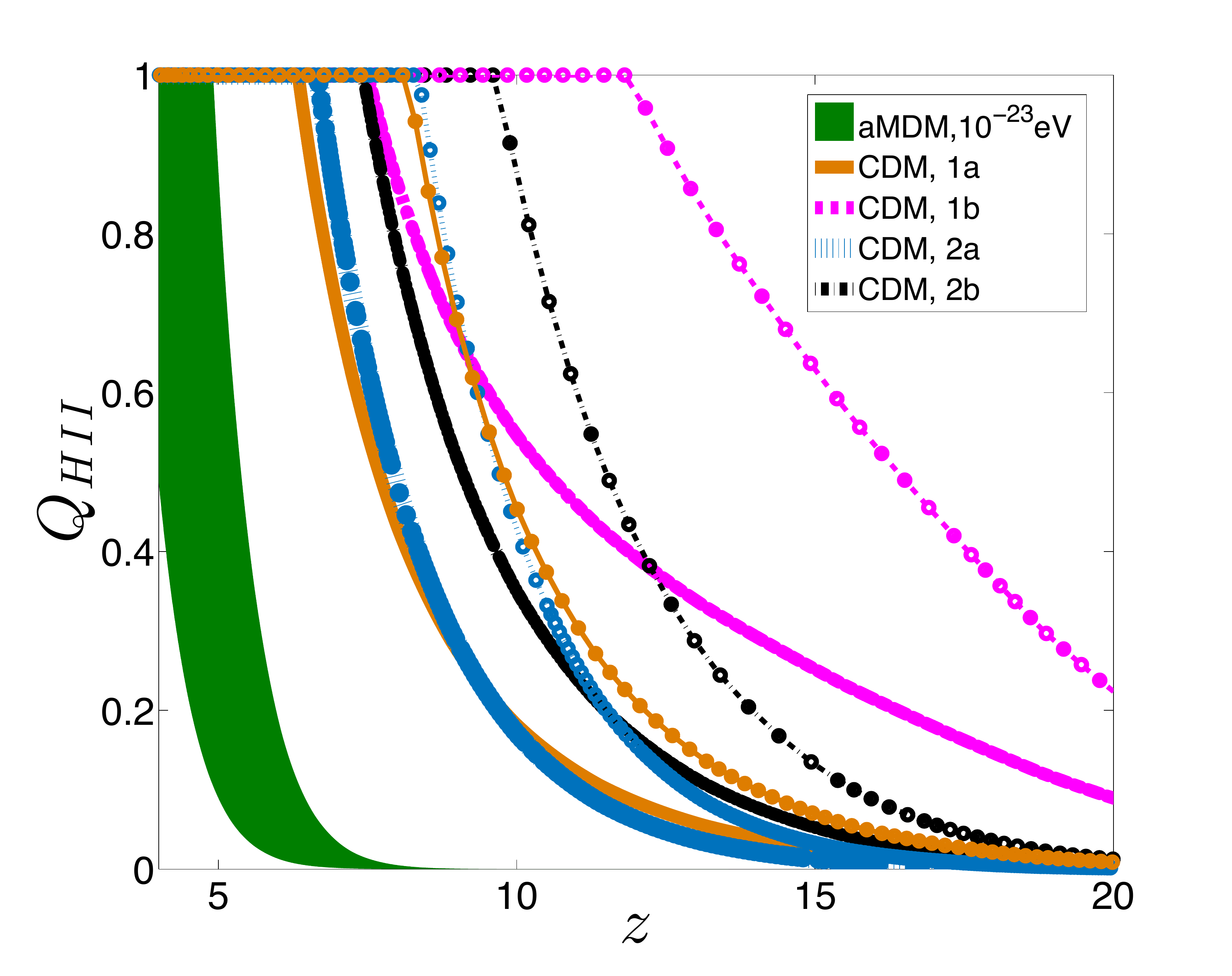} 
\includegraphics[width=0.49\textwidth]{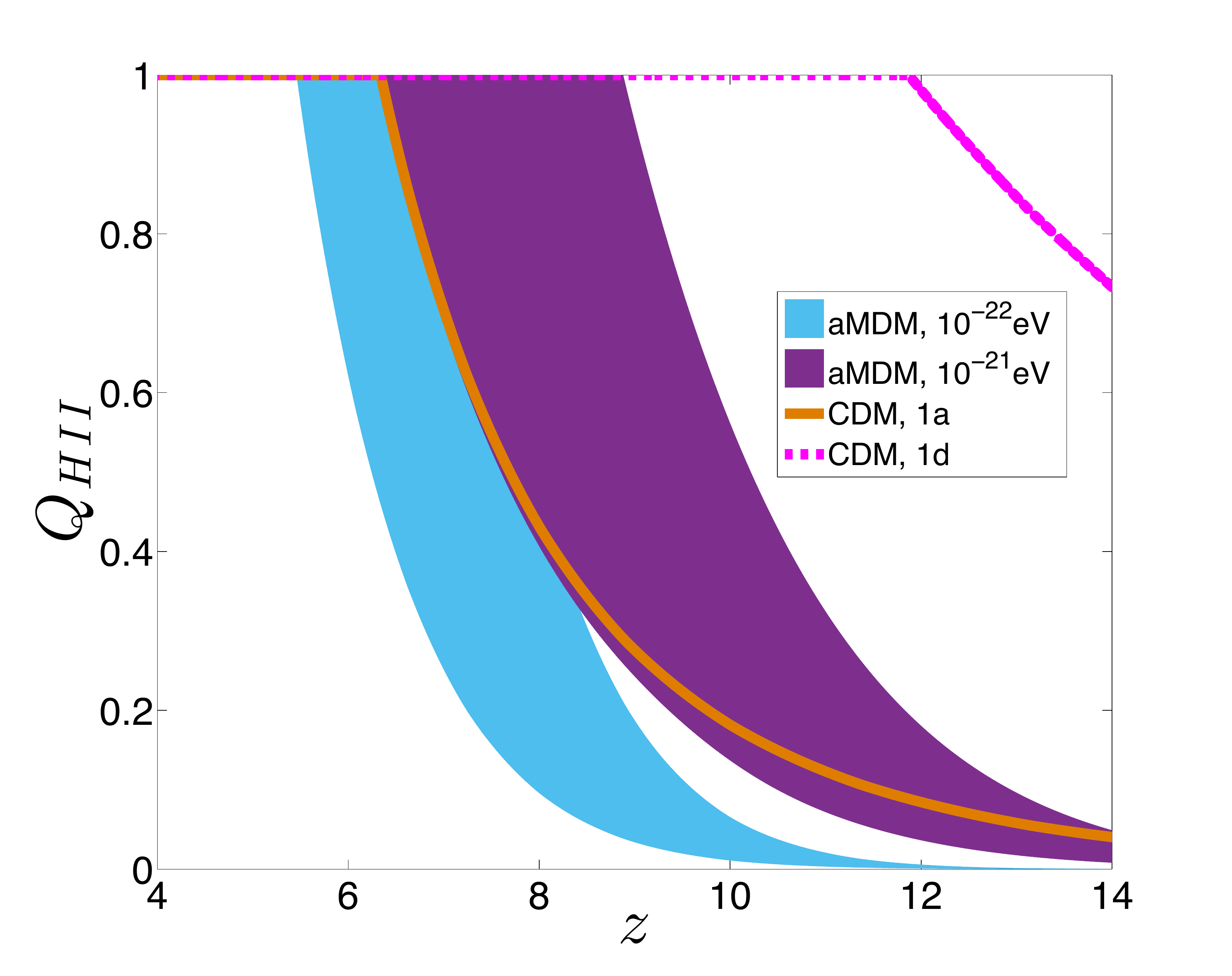} 
\caption{The reionization histories of the aMDM and CDM models. Left-hand panel: CDM `1a' (thick, solid orange), CDM `1b' (thick, dashed pink); CDM `1c' (thin, solid orange /circle markers); CDM `1d' (thin, dashed pink/circle markers); CDM `2a' (thick, dotted blue); CDM `2b' (thick, dash--dotted black); CDM `2c' (thin, dotted blue/circle markers); CDM `2d' (thin, dash--dotted black/circle markers). The full range of $m_a=10^{-23}\text{ eV}$ reionization histories are represented by the green patch. The $m_a=10^{-23}\text{ eV}$ aMDM models are unable to reionize the universe by $z=5$. Right-hand panel:
the full range of $m_a=10^{-21}\text{ eV}$ reionization histories are represented by the dark purple patch and the $m_a=10^{-22}\text{ eV}$ reionization histories are represented by the light blue patch. CDM models `1a' and `1d' are shown for reference (\textit{x}-axis has a different scale in each panel). The $m_a=10^{-22}\text{ eV}$ aMDM model reionization histories complete reionization by $z=6$ depending on the assumed reionization parameters. All models complete reionisaiton by $z=5.5$. The reionization histories of the $m_a=10^{-21}\text{ eV}$ aMDM models complete reionization by $z=6$.}
\label{fig:Q_a}
\end{center}
\end{figure*}

The reionization histories of the aMDM and CDM models, $Q_{\text{H\rom{2}}}(z)$, are shown in the left- and right-hand panels of Fig.~\ref{fig:Q_a}. The evolution of $Q_{\text{H\rom{2}}}(z)$ in a CDM cosmology, shown in the left-hand panel of Fig.~\ref{fig:Q_a}, depends strongly on the Schechter function model, the assumed value of the escape fraction of ionizing photons, $f_{\rm esc}$, and the minimum UV magnitude, $M_{\rm lim}$. The extreme reionization history of CDM model `1d' (dashed pink curve) is likely to be unphysical, but is included in Fig. \ref{fig:Q_a} in order to illustrate the full range of reionization histories produced by case `d' reionization models and to facilitate comparison with aMDM model `4d' results. 

The $m_a=10^{-23}\text{ eV}$ aMDM models with axions comprising $100$ per cent or $50$ per cent of the DM are unable to reionize the universe by $z=5$ in the wide range of reionization models we consider, as shown by the blue shaded region of the left-hand panel of Fig. \ref{fig:Q_a}. Observations of the Gunn--Peterson trough in quasar spectra at $z>6$ \citep{2006AJ....132..117F} and transmission in the Ly$\alpha$ forest for $z<6$ \citep{2001AJ....122.2850B, 2001ApJ...560L...5D, 2001AJ....122.2833F} suggests the epoch of reionization ends at $z \sim 6$ ($Q_{\text{H\rom{2}}}(z\sim6) > 0.99$). These constraints on the neutral fraction of hydrogen at the end of reionization rely on detailed modelling of both the IGM and ionizing sources making their accuracy (and therefore the exact end of reionization) the subject of debate \citep{2007ApJ...662...72B,2009MNRAS.394.1667F, 2010MNRAS.407.1328M, 2011MNRAS.415.3237M, 2013ApJ...768...71R}. \citet{2011MNRAS.415.3237M} use the covering fraction of `dark' pixels in quasar spectra to obtain the more conservative constraints of $Q_{\text{H\rom{2}}}(z=5.5) >0.8$ and $Q_{\text{H\rom{2}}}(z=6)> 0.5$ on the end of reionization. The most extreme $m_a=10^{-23}\text{ eV}$ model `4d' value of $Q_{\text{H\rom{2}}}(z=6) = 0.19$ is inconsistent with even these more conservative constraints. The reionization history therefore rules out $m_a=10^{-23}\text{ eV}$ from contributing more than half of the DM, consistent with the constraints of Section~\ref{sec:cum_lum}.

The $m_a=10^{-21}\text{ eV}$ aMDM models have a range of reionization histories, depicted by the dark purple (rightmost) shaded region in the right-hand panel of Fig.~\ref{fig:Q_a}, depending on the assumed value of the escape fraction and Schecter function fit. Each case completes reionization by $z=6$. The reionization histories of the most conservative reionization model, `1a', for $m_a=10^{-21}\text{ eV}$, represented by the leftmost edge of the dark purple shaded region, and CDM (orange, solid curve) are similar for $z<8$. The early reionization history ($Q_{\text{H\rom{2}}}(z>8)$), is more extended for CDM. The range of possible reionization histories is less varied for the $m_a=10^{-21}\text{ eV}$ aMDM models than CDM, as seen in the comparison of the right edge of the purple (right) shaded region with the pink (dashed) curve in Fig. \ref{fig:Q_a} (right-hand panel). This is due to the delay in the build-up of small-mass haloes in an $m_a=10^{-21}\text{ eV}$ aMDM cosmology compared to CDM such that the $m_a=10^{-21}\text{ eV}$ aMDM early reionization history is less affected by changes in reionization model assumptions. We see that it is possible for the reionization history of the Universe to distinguish CDM from $m_a=10^{-21}\text{ eV}$ under certain assumptions. We will return to this question in Section~\ref{sec:dur_reion}.

The reionization histories of the $m_a=10^{-22}\text{ eV}$ aMDM models
represented by the blue shaded (leftmost) region in Fig.~\ref{fig:Q_a} complete
reionization by $z=6$ depending on the assumed value of the escape
fraction, Schechter function fit, and the axion fraction of dark
matter. A larger escape fraction ($f_{\rm esc} = 0.5$) or a smaller
axion fraction of DM ($\Omega_{\rm a}/\Omega_{\rm d} = 0.5$) is required to
complete reionization by $z=6$. All models complete reionization by
$z=5.5$ and are consistent with the more conservative Ly$\alpha$
constraints of \citet{2011MNRAS.415.3237M}. The $m_a=10^{-22}\text{ eV}$ aMDM
model reionization histories are more abbreviated compared to CDM due
to the relative delay in small-mass galaxy formation. In most cases
for the $m_a=10^{-22}\text{ eV}$ aMDM models, changing the limiting
magnitude from $M_{\rm lim}=-10$ to $-13$ produces little
to no change in $Q_{\text{H\rom{2}}}(z)$ as both limits fall below the
magnitude where the $M_{\rm h}(M_{\rm UV})$ relation truncates. Due to the weaker dependence on model assumptions reionization is able to constrain $m_a=10^{-22}\text{ eV}$ rather well.

\begin{figure*}
\begin{center}
\includegraphics[width=0.49\textwidth]{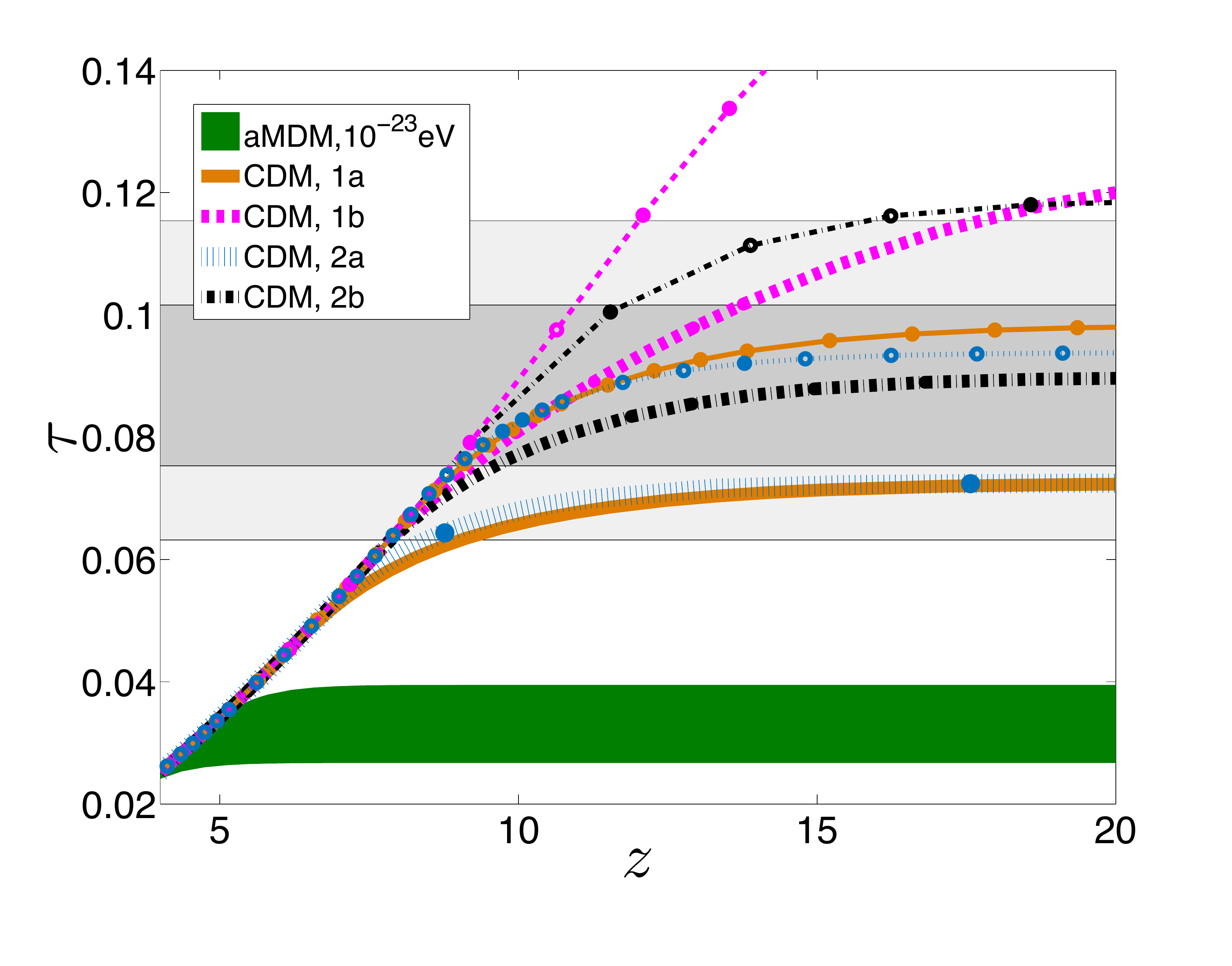} 
\includegraphics[width=0.49\textwidth]{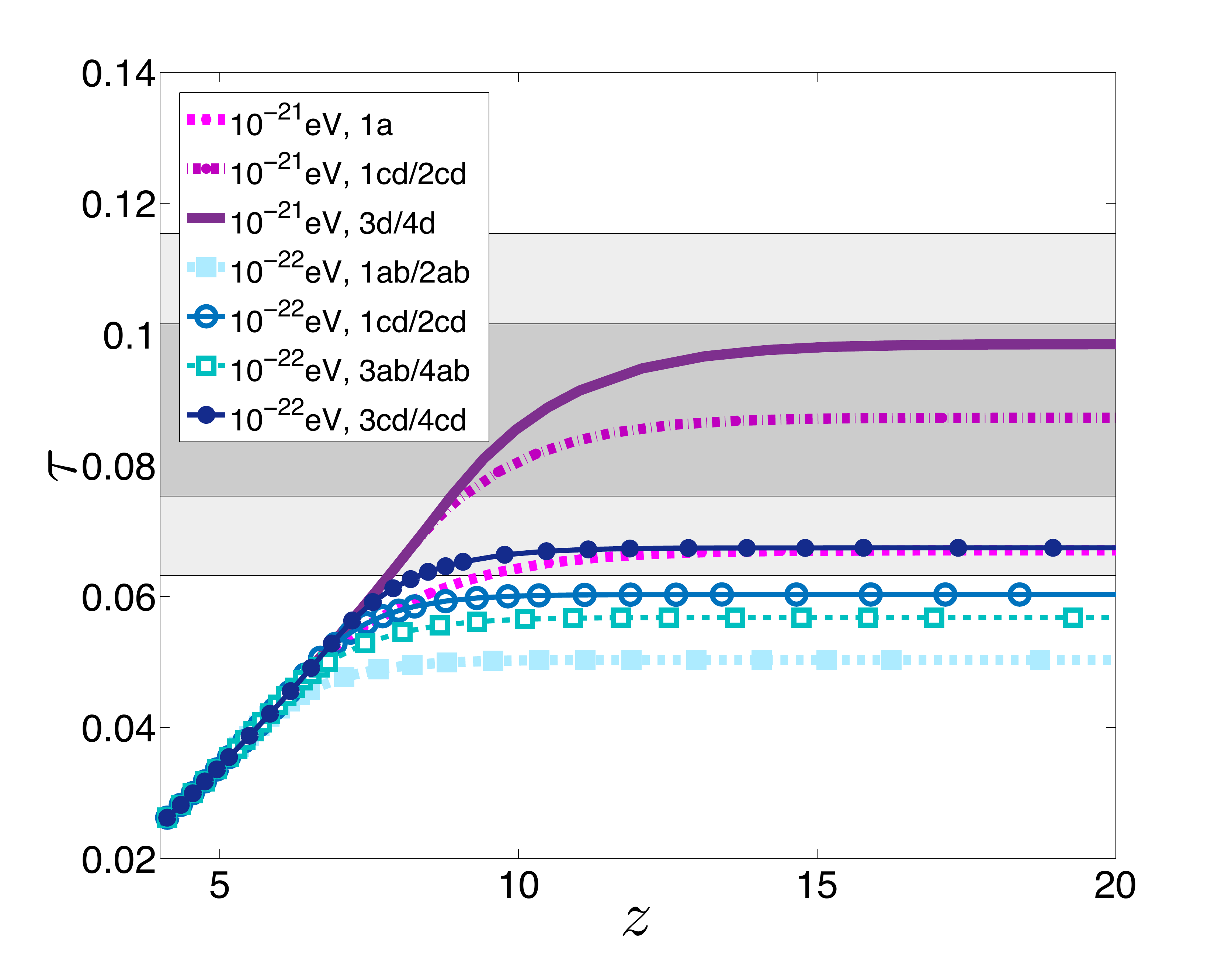} 
\caption{The CMB optical depth, $\tau$, for the aMDM and CDM models. The grey (horizontal) bands in both panels are \textit{Planck}+\textit{WMAP} $68$ per cent $(1\sigma$,dark grey) and $95.45$ per cent $(2\sigma$,light grey) confidence levels. Left-hand panel: CDM `1a' (thick, solid orange), CDM `1b' (thick, dashed pink); CDM `1c' (thin, solid orange/circle markers); CDM `1d' (thin, dashed pink/circle markers); CDM `2a' (thick, dotted blue); CDM `2b' (thick, dash--dotted black); CDM `2c' (thin, dotted blue/circle markers); CDM `2d' (thin, dash--dotted black/circle markers). The full range of $m_a=10^{-23}\text{ eV}$ CMB optical depth values are represented by the shaded green patch. The $m_a=10^{-23}\text{ eV}$ aMDM model is excluded at greater than $99.99$ per cent confidence. Right-hand panel: $m_a=10^{-22}\text{ eV}$ models `1ab/2ab' (dashed, light blue curve/filled square markers); $m_a=10^{-22}\text{ eV}$ models `1cd/2cd' (solid, blue/open circle); $m_a=10^{-22}\text{ eV}$ models `3ab/4ab' (dashed, cyan/open square); $m_a=10^{-22}\text{ eV}$ models `3cd/4cd' (solid, dark blue/filled circle); $m_a=10^{-21}\text{ eV}$ `1a' (dashed magenta); $m_a=10^{-21}\text{ eV}$ models `1cd/2cd' (dash--dotted purple); $m_a=10^{-21}\text{ eV}$ models `1cd/2cd' (solid dark purple). The $m_a=10^{-22}\text{ eV}$ aMDM model predictions for $\tau$ (right-hand panel) is in tension with \textit{Planck}+\textit{WMAP} constraints. Only the $m_a=10^{-22}\text{ eV}$ model with axions contributing only 50 per cent of the DM and with the most extreme reionization assumptions is consistent at $95.45$ per cent ($2\sigma$) confidence. Less conservative models with larger DM fraction in axions are in more tension. The $m_a=10^{-21}\text{ eV}$ aMDM model $\tau$ predictions are all consistent with \textit{Planck}+\textit{WMAP} constraints and depend strongly on the reionization parameter assumptions.}
\label{fig:tau_a}
\end{center}
\end{figure*}

\subsection{CMB optical depth}

The left- and right-hand panels of Fig.~\ref{fig:tau_a} show the predictions for the CMB optical depth, $\tau$, for the aMDM and CDM models. The CMB optical depth is plotted cumulatively as a function of redshift. The predicted full-integrated value of $\tau$ for each model can be taken from the high-redshift end of the plot for comparison with CMB measurements. The grey (horizontal) bands in both figures show the $68$ per cent $(1\sigma)$ and $95.45$ per cent $(2\sigma)$ confidence levels around the maximum likelihood value of $\tau = 0.0891$ from the recent \textit{Planck}+\textit{WMAP} analysis by \cite{spergel2013}. We will quote results for the $99.73$ per cent $(3\sigma)$ confidence level where applicable; however, this region is not plotted in either figure for simplicity. 

The CDM and $m_a=10^{-23}\text{ eV}$ aMDM model predictions for $\tau$ are shown in the left-hand panel of Fig. \ref{fig:tau_a}. The $\tau$ values predicted for the CDM models depend strongly on the model's reionization assumptions. The more conservative assumptions of CDM models `1a' and `1b' give results that are consistent with \textit{Planck}+\textit{WMAP} constraints at the $95.5$ per cent confidence level.  Only CDM model `1d' is ruled out at more than $99.99$ per cent confidence and is included here to contrast against the aMDM maximal reionization models. 

The green (lower) shaded patch in the left-hand panel of Fig.~\ref{fig:tau_a} shows the full range of possible CMB optical depth values for the $m_a=10^{-23}\text{ eV}$ aMDM model. The upper bound (given by model `4d') is excluded at greater than $99.73$ per cent confidence. The slow build-up of small galaxies in the $m_a=10^{-23}\text{ eV}$ aMDM models, which produces a delayed reionization history in this cosmology, prohibits the model from reproducing a value of $\tau$ consistent with \textit{Planck}+\textit{WMAP} constraints. We see yet again that, taking into account a wide range of models for reionization to bracket our systematic uncertainty, $m_a=10^{-23}$\text{ eV} contributing more than 50 per cent of the DM is ruled out, consistent with the constraints from Sections~\ref{sec:cum_lum} and \ref{sec:q_hist}.

The CMB optical depth predictions for the $m_a=10^{-21}\text{ eV}$ and the $m_a=10^{-22}\text{ eV}$ aMDM models are shown in the right-hand panel of Fig.~\ref{fig:tau_a}. The $m_a=10^{-22}\text{ eV}$ aMDM model predictions for $\tau$ are depicted by the four blue curves shown in Fig. \ref{fig:tau_a}. The predicted $\tau$ values of models `1a', `1b', `2a', `2b' are excluded by \textit{Planck}+\textit{WMAP} constraints at the $99.73$ per cent confidence level. Therefore, in conservative models of reionization, $m_a=10^{-22}\text{ eV}$ is excluded from being all of the DM at more than $3\sigma$. The other $m_a=10^{-22}\text{ eV}$ aMDM model predictions for $\tau$, however, are well within the $99.73$ per cent confidence region, and the $\tau$ values of models `3c', `3d', `4c', and `4d' are within the $95.5$ per cent confidence region. In the more extreme models of reionization, $m_a=10^{-22}\text{ eV}$ is allowed to contribute up to half of the DM while remaining consistent with the observed value of $\tau$ at $2\sigma$.

The Schechter function fit and the value of limiting magnitude have little or no effect on the predicted $\tau$ values for all axion fractions of DM of the $m_a=10^{-22}\text{ eV}$ aMDM model. The assumed value of the escape fraction of ionizing photons is the only reionization parameter that strongly affects the predicted value of $\tau$ for this aMDM model. Tighter constraints on the observed value of $\tau$ (with the same maximum likelihood value) by future CMB experiments could place considerable tension on this model. 

The upper and lower bounds of the $m_a=10^{-21}\text{ eV}$ aMDM model predictions for $\tau$ are, respectively, represented by the solid, dark purple curves (models `3d' and `4d') and dashed magenta curves (model `1a') in the right-hand panel of Fig. \ref{fig:tau_a}. CMB optical depth values predicted for all other $m_a=10^{-21}\text{ eV}$ aMDM models are within these two bounding curves and are consistent with \textit{Planck}+\textit{WMAP} constraints for all axion fractions of DM and reionization model assumptions. The $\tau$ prediction of models `1c', `1d', `2c', and `2d' that are closest to the maximum likelihood value of \textit{Planck}+\textit{WMAP} are represented by the dashed magenta curve in the right-hand panel of Fig. \ref{fig:tau_a}.

The reionization parameter assumptions have a larger effect on the $m_a=10^{-21}\text{ eV}$ aMDM model than the other aMDM models. The most important reionization parameter is the value of the escape fraction. The subsequent importance of the limiting magnitude and choice of Schechter function depends on the axion fraction of DM and the value of the escape fraction. For example, there is a small spread in the predicted value of of $\tau$ for models `1a', `1b', `2a', and `2b', while the models `1c', `1d', `2c', and `2d' are all very similar. For axion fractions of DM $\Omega_{\rm a}/\Omega_{\rm d} = 0.5$ (models `3' and `4'), the choice of limiting magnitude effects the predicted value of $\tau$, but the Schechter function fit does not, i.e. the $\tau$ values of models `3d' and `4d' are similar and greater than the similar $\tau$ values of  models `3c' and `4c' (not shown in Fig. \ref{fig:tau_a}). 

\subsection{Measuring the duration of reionization}
\label{sec:dur_reion}

\begin{figure*}
\begin{center}
\begin{tabular}{ll}
\includegraphics[width=0.49\textwidth]{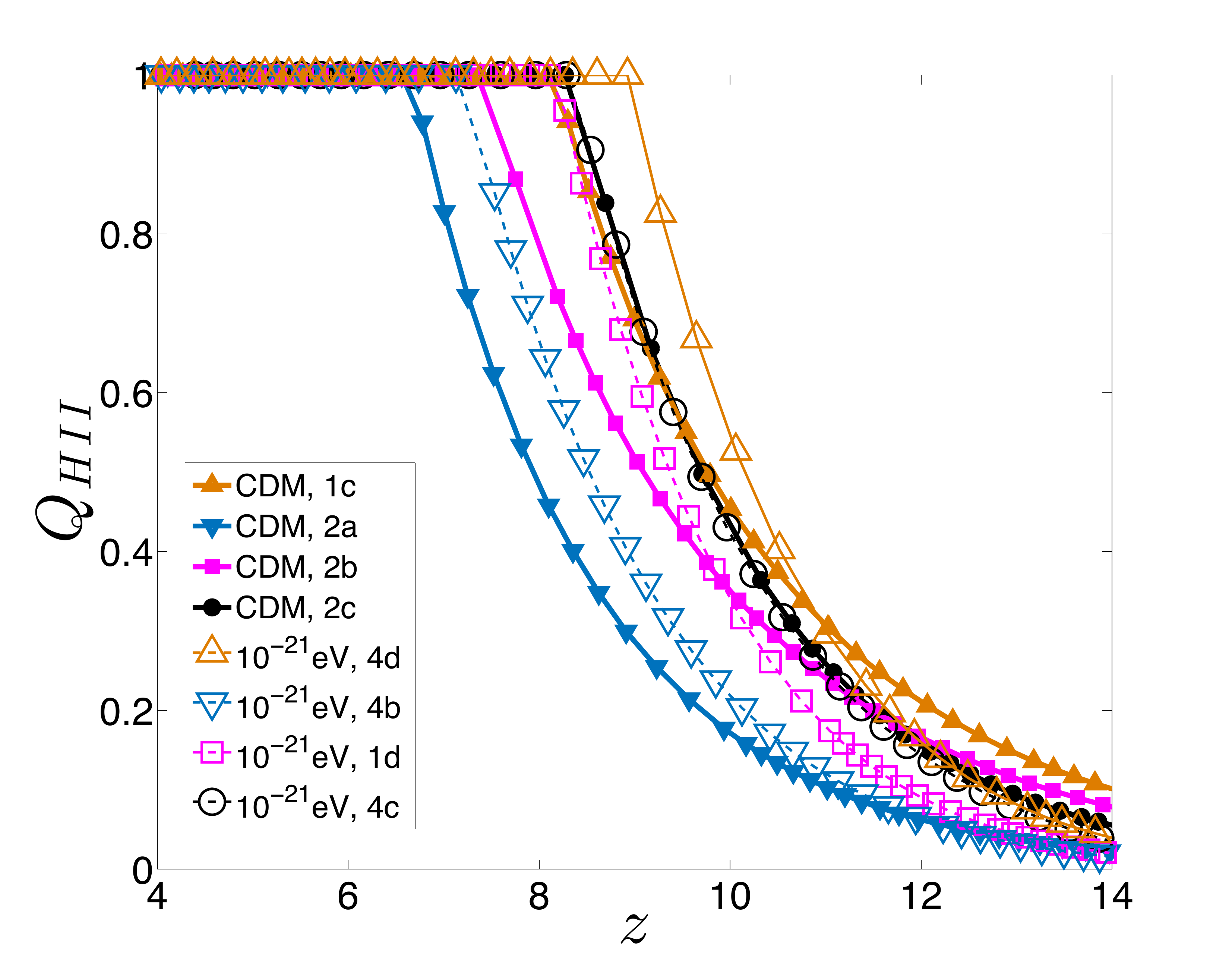} &
\includegraphics[width=0.49\textwidth]{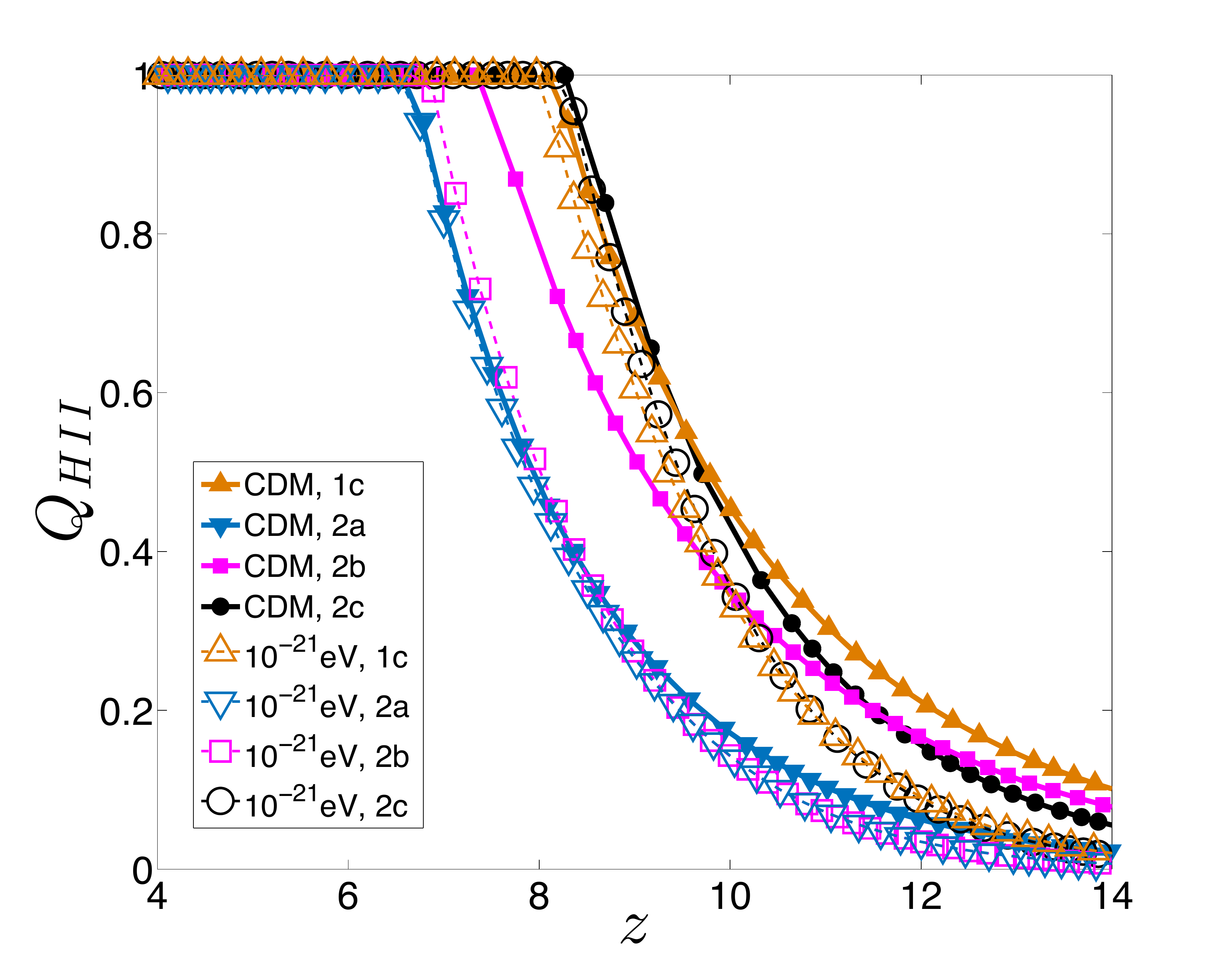} \\
\includegraphics[width=0.49\textwidth]{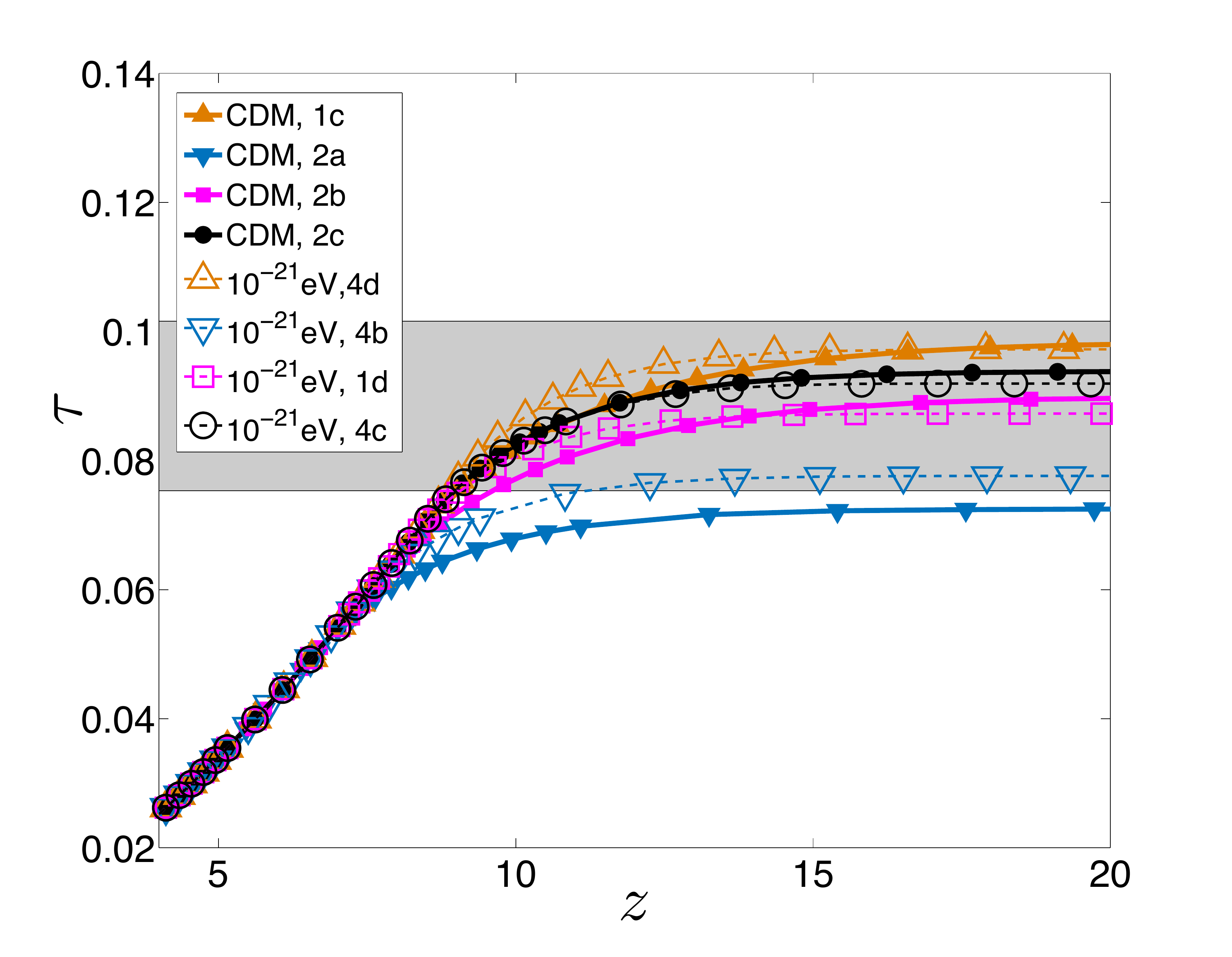} &
\includegraphics[width=0.49\textwidth]{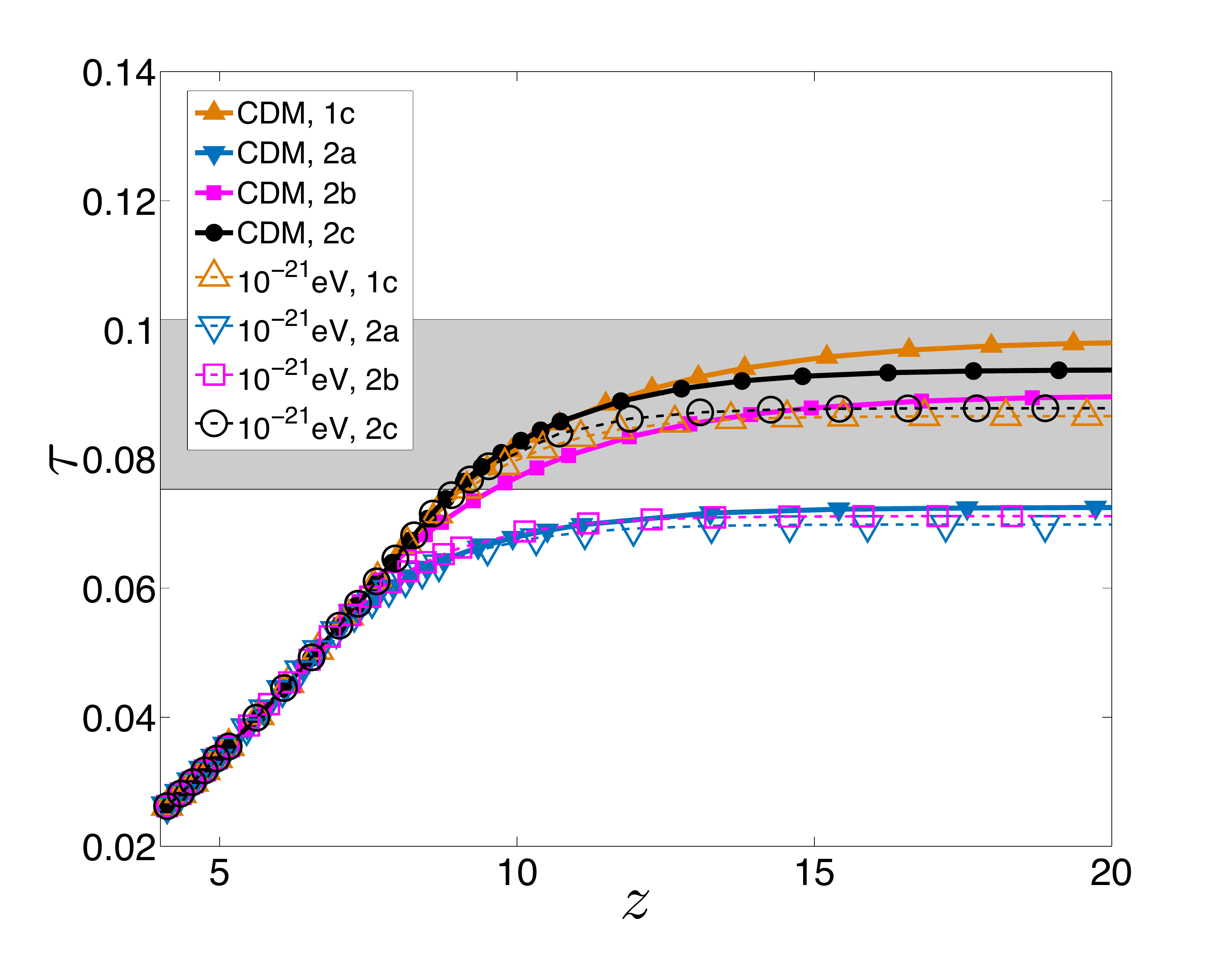} \\ 
\end{tabular}
\caption{The reionization histories (top row) and CMB optical depth values (bottom row) of two sets of four $m_a=10^{-21}\text{ eV}$ aMDM models compared to a set of four CDM models $\{$`1c' (solid orange curve/filled triangle marker), `2a' (solid blue/upside-down, filled triangle), `2b' (solid pink/filled square), `2c' (solid black/filled circle)$\}$. Left-hand panels: the first aMDM set $\{$ `4d' (dashed orange/open triangle), `4b' (dashed blue/open upside-down triangle), `1d' (dashed pink/open square), `4c' (dashed black/open circle)$\}$ is selected $\tau$ consistent at $1\sigma$ (grey band) with \textit{Planck}+\textit{WMAP} value and provide a close match to each CDM model's $\tau$ prediction. Right-hand panels: $\{$ `1c' (dashed orange/open triangle), `2a' (dashed blue/open upside-down triangle), `2b' (dashed pink/open square), `2c' (dashed black/open circle)$\}$ have the same reionization assumptions as the four CDM models. Models that are differentiable from CDM by the duration of reionization (according to AdvACT constraints): `Set 1' - (`1d' and `4d') and `Set 2' -- (`1c', `2b', and `2c')}
\label{fig:tau_comp}
\end{center}
\end{figure*}

Future small-scale CMB polarization measurements, such as the proposed Advanced ACTPol (AdvACT) experiment, aim to constrain the epoch of reionization through an accurate measurement of the kinematic Sunyaev--Zel'dovich (kSZ) effect \citep{1980MNRAS.190..413S} by breaking degeneracies between primary and secondary contributions to temperature anisotropies \citep{Calabrese:2014gwa}. The AdvACT experiment measurement of the patchy kSZ power spectrum amplitude and multipole shape could constrain the duration of reionization ($\delta z_{\rm re} = z_f(Q_{\text{H\rom{2}}}=0.75) - z_i(Q_{\text{H\rom{2}}}=0.25)$) and the median redshift of reionization $z_{\rm re,med} = z(Q_{\text{H\rom{2}}}=0.5)$ respectively, to an uncertainty of  $\sigma(\delta z_{\rm re}) = 0.2$ and $\sigma(z_{\rm re,med}) = 1.1$. 

\begin{table*}
\label{tab:Advact}
\begin{center} 
\caption{AdvACT constraints on $m_a=10^{-21}\text{ eV}$ aMDM models}
\begin{tabular}{l | r | r | l | r | r | r | r}
\hline
aMDM model & $\delta z_{\rm re}$ & $z_{\rm re,med}$ & CDM model & $\delta z_{\rm re}$ & $z_{\rm re,med}$ & CL$\{z_{\rm re}\}$ & CL$\{\delta z_{\rm re}\}$  \\ \hline
4d & 1.86 & 10.15 & Model 1c & 2.73 & 9.77 & 27.0 per cent & 99.9989 per cent  \\ 
4b & 1.99 & 8.52   & Model 2a & 2.08 & 7.93 & 40.8 per cent & 38.3 per cent  \\  
1d & 1.80 & 9.37   & Model 2b & 2.72 & 9.09 & 20.1 per cent & 99.9996 per cent  \\ 
4c & 2.10 & 9.68   &  Model 2c & 2.17 & 9.70 & 1.5 per cent & 31.1 per cent  \\ 
\hline
1c & 1.90 & 9.35   & Model 1c  & 2.73 & 9.77 & 29.7 per cent & 99.997 per cent \\ 
2a & 1.95 & 7.88   & Model 2a & 2.08 & 7.93 & 3.6 per cent & 51.6 per cent \\  
2b & 1.80 & 8.01   & Model 2b & 2.72 & 9.09 & 67.4 per cent & 99.997 per cent \\  
2c & 1.73 & 9.47   & Model 2c & 2.17 & 9.70 & 16.6 per cent & 97.2 per cent  \\  
\hline
\end{tabular}
\\
\end{center}
Column 1: aMDM model label, Column 2: aMDM model duration of reionization ($\delta z_{\rm re} = z_f(Q_{\text{H\rom{2}}}=0.75) - z_i(Q=0.25)$), Column 3: aMDM model median redshift of reionization $z_{\rm re,med} = z(Q_{\text{H\rom{2}}}=0.5)$, Column 4: CDM model label, Column 5: CDM model duration of reionization, Column 6: CDM model median redshift of reionization, Column 7: confidence level of AdvACT ability to differentiate between aMDM and CDM model's median redshift of reionization, Column 8: confidence level of AdvACT ability to differentiate between aMDM and CDM model's duration of reionization. The set of four aMDM models above the horizontal line make up `Set 1' and the four aMDM models below the horizontal line make up `Set 2'. The aMDM models that are distinguishable from CDM based on AdvACT constraints on the duration of reionization: `Set 1' - (`Model 1d' and `Model 4d'), `Set 2' - (`Model 1c', `Model 2b', and `Model 2c').
\end{table*}

We explore the ability of the AdvACT experiment to distinguish between the $m_a=10^{-21}\text{ eV}$ aMDM and CDM models by examining the AdvACT constraints on a set of reionization models for CDM and aMDM.  We chose four CDM models that span a range of reionization assumptions and predict the corresponding  $\tau$
values that are consistent with \textit{Planck}+\textit{WMAP} at $2\sigma$, namely
`1c', `2a', `2b', and `2c'. We compare
two sets of $m_a=10^{-21}\text{ eV}$ aMDM models to this set of CDM
models. The first set ( `4d', `4b', `1d', and
`4c'), shown in the left-hand panels of
Fig. \ref{fig:tau_comp}, is selected to have a predicted value of $\tau$ that is consistent with \textit{Planck}+\textit{WMAP} at $1\sigma$ and as close a match to the selected CDM model's $\tau$ prediction as possible. The second set, shown in the right-hand panels of Fig. \ref{fig:tau_comp}, is selected to have the same
reionization assumptions as the four CDM models and an axion fraction of DM of $\Omega_{\rm a}/\Omega_{\rm d} = 1.0$. The duration of reionization, the median redshift of reionization, and the
discriminating power of AdvACT to separate comparable models are listed
in Table 2.

Two of the four models in the first set (`1d' and `4d'), shown in the left-hand column of Fig. \ref{fig:tau_comp}, have a
duration of reionization that is differentiable from their CDM
counterpart at $4\sigma$, even though their median redshift of
reionization is within $1\sigma$ of CDM and they share a similar prediction for
$\tau$. The more extreme `d' models of reionization, with large escape fraction and limiting magnitude, are necessary for $m_a=10^{-21}\text{ eV}$ to match $\tau$ values of less extreme CDM models. Yet these models complete reionization more rapidly than their CDM counterparts and can thus be distinguished AdvACT.

From the second set, three of the four $m_a=10^{-21}\text{ eV}$
aMDM models (`1c', `2b', and `2c'), have a reionization duration that is distinct from CDM at greater than $2\sigma$ under the same reionization assumptions. With the exception of the most conservative `a' model of reionization, the large axion fraction of DM leads the axion models to different values of $\tau$ from CDM, though all easily within $2\sigma$ of the \textit{Planck}+\textit{WMAP} constraint. Reionization again completes more rapidly with axion DM than CDM, which will allow AdvACT to constrain aMDM models with large fractions of axion DM.

Taken together, the results of these two sets indicate that reionization histories for a subset of $m_a=10^{-21}\text{ eV}$ aMDM models can be distinguished from CDM by AdvACT constraints. As discussed in Section \ref{sec:q_hist}, the $m_a=10^{-21}\text{ eV}$ aMDM and CDM models with more conservative reionization assumptions have similar reionization histories, while the CDM model has a more varied response to more extreme reionization assumptions. The median redshift of reionization, $z_{\rm re,med}$, for the CDM and aMDM models (listed in Table 2) provides a dividing line for CDM and aMDM models that can be differentiated by AdvACT and those that cannot. The aMDM and CDM models where both models have a median redshift of resionisation of $z_{\rm re,med} < 8.6$ are indistinguishable by AdvACT constraints. These aMDM and CDM models have more conservative reionization assumptions that give similar reionization histories, as illustrated by the aMDM models `4b' and `2a' compared with CDM model `2a' represented by the blue curves in Fig. \ref{fig:tau_comp}. If both aMDM and CDM models have a median redshift of reionization of $z_{\rm re,med} > 8.6$ they will have a duration of reionization that is differentiable by AdvACT constraints, if the aMDM model does not have the same reionization assumptions as the CDM model and an axion fraction of DM of $\Omega_{\rm a}/\Omega_{\rm d} = 0.5$. The black curves in Fig. \ref{fig:tau_comp} illustrate this last point. The aMDM model `4c' has an axion fraction of DM of $\Omega_{\rm a}/\Omega_{\rm d} = 0.5$ and has a reionization history that is indistinguishable from CDM model `2c' as shown by the black curves in the left-hand panels of Fig. \ref{fig:tau_comp}. The aMDM model `2c' has the same reionization assumptions as model `4c' and the CDM model `2c', but with an axion fraction of DM of $\Omega_{\rm a}/\Omega_{\rm d} = 1.0$ it has reionization history that is differentiable from CDM model `2c' as shown by the black curves in the right-hand panels of Fig. \ref{fig:tau_comp}.

Complementary data and analyses that could constrain the escape fraction of
ionizing photons, tighten the constraints on $\tau$, and place either
theoretical or observational constraints on the limiting magnitude of
the UV-luminosity function could improve on the range of
$m_a=10^{-21}\text{ eV}$ aMDM and CDM models that can be ruled out by
future CMB measurements. The proposed AdvACT experiment could improve
on constraints on reionization duration, median reionization
redshift, and $\tau$ value by extending the temperature fluctuation multipole space down to $l = 10$.

\section{Summary and Discussion} 
\label{sec:Summary}

We have used the model of \cite{marsh2013b} for the HMF
of ultralight aMDM \citep[which appears broadly
consistent with the simulations of ][]{schive2014} to place
constraints on the axion mass. To do this, we used the predicted high-$z$ UV-luminosity function compared to that derived from deep imaging with the \textit{HST} together with the predicted reionization
history of the universe, via the optical depth, $\tau$, compared to the value derived
from the CMB. For simplicity, we have considered only models where axions comprise either all or half of the total DM. We assume that galaxies are only contributor to reionization. AGN feedback could possibly loosen some constraints. Such a model, however, is severely contained by the diffuse X-ray background \citep{2004ApJ...613..646D}. 

We have found that both the UV-luminosity function and the optical depth consistently forbid $m_a=10^{-23}\text{ eV}$ from contributing a large fraction of the DM. This appears to exclude the possibility to search for ULAs via pulsar timing experiments as proposed by \cite{khmelnitsky2013}. We have found that $m_a=10^{-23}\text{ eV}$ cannot produce enough 
galaxies of the required magnitude at high-$z$ to be consistent with
HUDF. Under a wide range of limiting magnitudes and escape fractions,
allowing for a large uncertainty in the model for reionization,
$m_a=10^{-23}\text{ eV}$ fails to reionize the universe by $z=6$ and
is inconsistent with the measured value of $\tau$ at $>3\sigma$. In
terms of DM fraction with $m_a=10^{-23}\text{ eV}$, we have ruled out
both $100$ and $50$ per cent at greater than $8\sigma$. The strength of the constraints suggests that a more detailed study varying the fraction of DM in axions over a wider range will be able to limit the fraction still further at this mass. By simple extrapolation the entire ULA mass
range $10^{-32}\text{ eV}\lesssim m_a\lesssim 10^{-23} \text{ eV}$ is
excluded, by an order of magnitude, for $\Omega_{\rm a}/\Omega_{\rm d}\gtrsim 0.5$ fraction of the DM.\footnote{At the low-mass end, $m_a\lesssim
10^{-32}\text{ eV}$, axions behave quintessence and our constraints do
not apply (Marsh et al., in preparation).}

With $m_a=10^{-22}\text{ eV}$ it is just possible to produce enough high-$z$ galaxies to be consistent with HUDF. However, under conservative assumptions for the model of reionization, $m_a=10^{-22}\text{ eV}$ is in tension with the measured value of $\tau$ at $3\sigma$ if the DM is entirely composed of axions. It is interesting to note that this encompasses the best-fitting value of $m_a=8.1\times 10^{-23}\text{ eV}$ of \cite{schive2014} required to account for a core in Fornax \citep[see also][]{schive2014b}. This is a major result of this paper as it puts an important part of axion parameter space, which is related to the solution of the small-scale problems, under pressure. The tension from $\tau$ is reduced to $2\sigma$ if the axion is only 50 per cent of the DM, but then core formation is likely spoiled \citep{marsh2013b}. At high redshift, the central overdensity of axion DM haloes is decreased \citep{schive2014} compared to pure CDM haloes and the shallower potential well of the cored DM halo may impact galaxy formation in the early universe. While beyond the scope of this paper, a full investigation of the role of cored DM haloes on the star formation history, feedback efficiency, and the escape fraction of ionizing photons would provide valuable insight into galaxy formation in an aMDM cosmology. More detailed studies are needed in this area to determine whether ULAs can resolve the cusp-core problem while being consistent with the reionization history of the universe. 

We have found that $m_a=10^{-21}\text{ eV}$ is consistent with the high-$z$ UV-luminosity function and the reionization history of the universe under a wide range of models for abundance matching and reionization, and with current observations is indistinguishable from CDM. 

Can we push constraints on $m_a$ further, and what are the observational and theoretical motivations for doing so? 

A tensor-mode interpretation of primordial degree scale CMB $B$-mode
polarization, which may have been observed by BICEP2
\citep{Ade:2014xna}, forbids the entire range of $10^{-28}\text{
eV}\lesssim m_a \lesssim 10^{-18}\text{ eV}$ from contributing any
significant amount of the DM \citep{Marsh:2014qoa}.\footnote{For
larger masses than this it is possible for a ULA to contribute
significantly to the DM density while having a small decay constant and avoiding isocurvature constraints.} Any evidence for
the existence of a ULA in this mass range would therefore be a signal of
non-trivial axion dynamics during or after inflation
\citep[e.g.][]{conlon2008}, or a non-tensor source of $B$-modes
\citep[e.g.][]{pospelov2008}. Furthermore, a range of axion masses
$10^{-19}\text{ eV}\lesssim m_a \lesssim 10^{-18}\text{ eV}$ is
potentially ruled out from observations of spinning supermassive
black holes, due to the super-radiant instability that would otherwise
be present \citep[Pani, private communication]{arvanitaki2010,Pani:2012vp}. Improving cosmological constraints
on $m_a$ by an order of magnitude or more can thus close a remaining
gap in ULA parameter space, confirm or refute the role of axions in resolving the small-scale crises of CDM, and be of relevance to inflationary model building.

With $m_a=10^{-22}\text{ eV}$ the UV-luminosity function has no
support for $M_{\rm UV}\gtrsim -17$ at $z\geq 10$. The planned deep
field measurement of the luminosity function by \textit{JWST}, which we forecast to reach $M_{\rm UV}\approx -16$ at $z\geq 10$, could therefore
easily rule out, or find evidence for, this model. Furthermore, \cite{Calabrese:2014gwa}
showed that near-future improvements in the measurement of CMB
polarization by AdvACT will significantly
improve our knowledge of the epoch, $z_{\rm re}$, and duration,
$\Delta z_{\rm re}$, of reionization. Achieving $\sigma (z_{\rm
re})=1.1$ and $\sigma (\Delta z_{\rm re})=0.2$ could distinguish
$m_a=10^{-21}\text{ eV}$ from CDM, and also constrain the model of
reionization. Achieving these limits from \textit{JWST} and AdvACT on the axion contribution to DM for $m_a\gtrsim 10^{-22}\text{ eV}$
would be highly significant for ULA models of structure formation
\citep{marsh2013b,beyer2014, schive2014}, and for the parameter space
of the `string axiverse' \citep{axiverse2009}. As observational probes improve it is necessary to  study structure formation with axion DM further through theory and simulation, to keep up with the accuracy of the data.

\section*{Acknowledgements}

We acknowledge Ren\'{e}e Hlozek for supplying the $\tau$ likelihood and thank Marc Kamionkowski, Russell Ryan, Pierre Sikivie, Michael Strauss, and James Taylor for useful discussions. BB, JS and RFGW acknowledge support at JHU by NSF grant OIA-1124403. DJEM's research at Perimeter Institute is supported by the Government of Canada through Industry Canada and by the Province of Ontario through the Ministry of Research and Innovation. RFGW thanks the Aspen Center for Physics and NSF Grant PHY-1066293 for hospitality during the writing of this paper. The research of JS has been supported at IAP by the ERC project 267117 (DARK) hosted by Universit\'e Pierre et Marie Curie -Paris 6. 

\emph{Note added in proof:} Since the submission of this manuscript the Planck 2015 polarization results have been released \citep{planck_2015_params}, which give a slightly lower value of $\tau\approx 0.07--0.08\pm 0.002 $ (the central value depending on data set combinations). This lower value of $\tau$ no longer strongly disfavours $m_a=10^{-22}\text{ eV}$, allowing for a ULA solution to the cusp-core problem that, as shown here, makes additional, testable, predictions for future measurements of the epoch of reionization. This is discussed further by \cite{marsh_and_pop}.

\bibliography{axion_highz,doddy_new}
\bibliographystyle{mn2e.bst}

\end{document}